\documentclass[sigconf]{acmart}
\usepackage{algorithm}
\usepackage{algorithmic}
\usepackage{caption}
\usepackage{subfigure}
\usepackage{booktabs}
\usepackage{amsmath}
\usepackage{booktabs}

\usepackage{amsthm,amssymb}
\usepackage{balance}

\usepackage{amsmath}       
\usepackage{amsfonts}
\usepackage{bm}
\usepackage{graphicx}
\usepackage{multirow}
\usepackage{booktabs}
\usepackage{newfloat}
\usepackage{listings}
\usepackage{multirow}
\usepackage{subfigure}
\usepackage{multirow}
\usepackage{pdfx}
\usepackage{xcolor}         
\usepackage{makecell}
\newtheorem{thm}{Theorem}

\newtheorem{lem}{Lemma}

\newcommand{\myparatight}[1]{\smallskip\noindent{\bf {#1}:}~}

\AtBeginDocument{%
  \providecommand\BibTeX{{%
    \normalfont B\kern-0.5em{\scshape i\kern-0.25em b}\kern-0.8em\TeX}}}

\copyrightyear{2022}
\acmYear{2022}
\setcopyright{acmcopyright}\acmConference[KDD '22]{Proceedings of the 28th ACM SIGKDD Conference on Knowledge Discovery and Data Mining}{August 14--18, 2022}{Washington, DC, USA}
\acmBooktitle{Proceedings of the 28th ACM SIGKDD Conference on Knowledge Discovery and Data Mining (KDD '22), August 14--18, 2022, Washington, DC, USA}
\acmPrice{15.00}
\acmDOI{10.1145/3534678.3539231}
\acmISBN{978-1-4503-9385-0/22/08}

\begin{document}
\title{FLDetector: Defending Federated Learning Against Model Poisoning Attacks via Detecting Malicious Clients} 


\author{Zaixi Zhang}
\affiliation{%
  \country{University of Science and Technology of China}}
\email{zaixi@mail.ustc.edu.cn}

\author{Xiaoyu Cao}
\affiliation{%
  \country{Duke University}}
\email{xiaoyu.cao@duke.edu}

\author{Jinyuan Jia}
\affiliation{%
  \country{Duke University}}
\email{jinyuan.jia@duke.edu}

\author{Neil Zhenqiang Gong}
\affiliation{%
  \country{Duke University}}
\email{neil.gong@duke.edu}
\renewcommand{\shortauthors}{Zaixi Zhang et al.}


\begin{abstract}
Federated learning (FL) is vulnerable to \emph{model poisoning attacks}, in which malicious clients corrupt the global model via sending manipulated model updates to the server. Existing defenses mainly rely on Byzantine-robust or provably robust FL methods, which aim to learn an accurate global model even if some clients are malicious. 
However, they can only resist a \emph{small} number of malicious clients. It is still an open challenge how to defend against model poisoning attacks with a large number of malicious clients. Our FLDetector addresses this challenge via \emph{detecting} malicious clients.  FLDetector aims to detect and remove majority of the malicious clients  such that a Byzantine-robust or provably robust FL method can learn an accurate global model using the remaining clients. 
Our key observation is that, in model poisoning attacks, 
the model updates from a client in multiple iterations are inconsistent.  Therefore, FLDetector detects malicious clients via checking their model-updates consistency. Roughly speaking, the server predicts a client's model update in each iteration based on historical model updates, and flags a client as malicious if the received model update from the client and the predicted model update are inconsistent in multiple iterations. Our extensive experiments on three benchmark datasets show that FLDetector can accurately detect malicious clients in multiple state-of-the-art model poisoning attacks and adaptive attacks tailored to FLDetector. After removing the detected malicious clients, existing Byzantine-robust FL methods can learn accurate global models. 
\end{abstract}

\ccsdesc[500]{Security and privacy~Intrusion/anomaly detection and malware mitigation}
\ccsdesc[500]{Computing methodologies~Distributed artificial intelligence}

\keywords{Federated Learning; Model Poisoning Attack; Malicious Client Detection; Anomaly Detection}

\maketitle
\section{Introduction}
Federated Learning (FL) \cite{mcmahan2017communication, yang2019federated} is an emerging  learning paradigm over decentralized data. 
Specifically, multiple clients (e.g., smartphones, IoT devices, edge data centers) jointly learn a machine learning model (called \emph{global model}) without sharing their local training data with a cloud server. Roughly speaking, FL iteratively performs the following three steps: the server  sends the current gloabl model to the selected clients; each selected client  finetunes the received global model on its local training data and sends the  model update back to the server; the server aggregates the received model updates according to some aggregation rule and updates the global model.\par

However, due to its distributed nature, FL is vulnerable to \emph{model poisoning attacks}~\cite{fang2020local,bagdasaryan2020backdoor,baruch2019little,xie2019dba, bhagoji2019analyzing,cao2022mpaf}, in which the attacker-controlled malicious clients corrupt the global model via sending manipulated model updates to the server. The attacker-controlled malicious clients can be injected fake clients~\cite{cao2022mpaf} or genuine clients compromised by the attacker~\cite{fang2020local,bagdasaryan2020backdoor,baruch2019little,xie2019dba, bhagoji2019analyzing}. Based on the attack goals,  model poisoning attacks can be generally classified into \emph{untargeted} and \emph{targeted}. In the untargeted model poisoning attacks \cite{fang2020local,cao2022mpaf}, the corrupted global model indiscriminately makes incorrect predictions for a large number of testing  inputs. In the  targeted model poisoning attacks \cite{bagdasaryan2020backdoor,baruch2019little,xie2019dba,bhagoji2019analyzing}, the corrupted global model makes attacker-chosen, incorrect predictions for attacker-chosen testing inputs, while the global model's accuracy on other testing inputs is unaffected. For instance, the attacker-chosen testing inputs could be testing inputs embedded with an attacker-chosen trigger, which are also known as \emph{backdoor attacks}.

Existing defenses against model poisoning attacks mainly rely on Byzantine-robust FL methods~\cite{blanchard2017machine, yin2018byzantine, cao2020fltrust, chen2017distributed} (e.g., Krum \cite{blanchard2017machine} and FLTrust~\cite{cao2020fltrust}) or provably robust FL methods~\cite{cao2021provably}. These methods aim to learn an accurate global model even if some clients are malicious and send arbitrary model updates to the server. 
Byzantine-robust FL methods can theoretically bound the change of the global model parameters caused by malicious clients, while provably robust FL methods can guarantee a lower bound of testing accuracy under malicious clients. However, they are only robust to a small number of malicious clients~\cite{blanchard2017machine,yin2018byzantine,cao2021provably} or require a clean, representative validation dataset on the server~\cite{cao2020fltrust}. For instance, Krum can theoretically tolerate at most $\lfloor \frac{n-2}{2} \rfloor$ malicious clients.  FLTrust~\cite{cao2020fltrust} is robust against a large number of malicious clients but it requires the server to have access to a clean validation dataset whose distribution does not diverge too much from the overall training data distribution. As a result, in a typical FL scenario where the server does not have such a validation dataset, the global model can still be corrupted by a large number of malicious clients.

Li et al.~\cite{li2020learning} tried to detect malicious clients in model poisoning attacks. Their key assumption is that the model updates from malicious clients are statistically distinguishable with those from benign clients.  
In particular, they proposed to use a variational autoencoder (VAE) to capture model-updates statistics. Specifically, VAE assumes the server has access to a clean validation dataset that is from the overall training data distribution. Then, the server trains a model using the clean validation dataset. The model updates obtained during this process are used to train a VAE, which takes a model update as input and outputs a reconstructed model update. Finally, the server uses the trained VAE to detect malicious clients in FL. Specifically, if a client's model updates lead to high reconstruction errors in the VAE, then the server flags the client as malicious. However, this detection method suffers from two key limitations: 1) it requires the server to have access to a clean validation dataset, and 2) it is ineffective when the malicious clients and benign clients have statistically indistinguishable model updates.

In this work, we propose a new malicious-client detection method called \emph{FLDetector}. First, FLDetector addresses the limitations of existing detection methods such as the requirement of clean validation datasets. Moreover, FLDetector can be combined with Byzantine-robust FL methods, i.e., after FLDetector detects and removes majority of the malicious clients, Byzantine-robust FL methods can learn accurate global models. 
Our key intuition is that, benign clients calculate their model updates based on the FL algorithm and their local training data, while malicious clients craft the model updates instead of following the FL algorithm. As a result, the model updates from a malicious client are \emph{inconsistent} in different iterations. Based on the intuition,  FLDetector detects malicious clients via checking their model-updates consistency. 

Specifically, we propose that the server predicts  each client's model update in each iteration  based on historical model updates using the Cauchy mean value theorem. 
Our predicted model update for a client is similar to the client's actual model update if the client follows the  FL algorithm. In other words, our predicted model update for a benign (or malicious)
 client is similar (or dissimilar) to the model update that the client sends to the server. We use Euclidean distance to measure the similarity between a predicted model update and the received model update for each client in each iteration. Moreover, we define a \emph{suspicious score} for each client, which is dynamically updated in each iteration. Specifically, a client's suspicious score in iteration $t$ is the average of such Euclidean distances in the previous $N$ iterations. Finally, we leverage  $k$-means with Gap statistics based on the clients' suspicious scores to detect  malicious clients in each iteration. In particular, if the clients can be grouped into more than one cluster based on the suspicious scores and Gap statistics in a certain iteration,  we group the clients into two clusters using $k$-means and classify the clients in the cluster with larger average suspicious scores as malicious. 
 
We evaluate FLDetector on three benchmark datasets as well as one untargeted model poisoning attack~\cite{fang2020local}, three targeted model poisoning attacks~\cite{bagdasaryan2020backdoor,baruch2019little,xie2019dba}, and adaptive attacks tailored to FLDetector. Our results show that, for the untargeted model poisoning attack, FLDetector outperforms the baseline detection methods;  for the targeted model poisoning attacks, FLDetector outperforms the baseline detection methods in most cases and achieves comparable detection accuracy in the remaining cases; and FLDetector is effective against adaptive attacks. Moreover, even if FLDetector misses a small fraction of malicious clients,  after removing the  clients detected as malicious, Byzantine-robust FL methods can learn as accurate global models  as  when there are no malicious clients.

In summary, we  make the following contributions.
\begin{itemize}
    \item We perform a systematic study on defending FL against model poisoning attacks via detecting malicious clients. 
    
    \item We propose FLDetector, an unsupervised method, to detect malicious clients via checking the consistency between the received and predicted model updates of clients.

    \item We empirically evaluate FLDetector against multiple state-of-the-art model poisoning attacks and adaptive attacks on three benchmark datasets. 
    
\end{itemize}

\section{Related Work}
\subsection{Model Poisoning Attacks against FL}
Model poisoning attacks generally can be untargeted~\cite{fang2020local, cao2022mpaf, shejwalkar2021manipulating} and  targeted~\cite{bagdasaryan2020backdoor,baruch2019little,xie2019dba,bhagoji2019analyzing}. Below, we review one state-of-the-art untargeted attack and three targeted attacks.

\myparatight{Untargeted Model Poisoning Attack}  Untargeted model poisoning attacks aim to corrupt the global model such that it has a low accuracy for indiscriminate testing inputs. Fang et al. \cite{fang2020local} proposed an untargeted attack framework against FL. Generally speaking, the framework formulates  untargeted attack as an optimization problem, whose solutions are the optimal crafted model updates on the malicious clients that maximize the difference between the aggregated model updates before and after the attack. The framework can be applied to any aggregation rule, e.g., they have  shown that the framework can substantially reduce the testing accuracy of the global models learnt by FedAvg~\cite{mcmahan2017communication}, Krum~\cite{blanchard2017machine}, Trimmed-Mean~\cite{yin2018byzantine}, and Median~\cite{yin2018byzantine}.

\myparatight{Scaling Attack, Distributed Backdoor Attack, and A Little is Enough Attack} In these targeted model poisoning attacks (also known as backdoor attacks), the corrupted global model predicts an attacker-chosen label for any testing input embedded with an attacker-chosen trigger. For instance, the trigger could be a patch located at the bottom right corner of an input image. 
Specifically, in Scaling Attack \cite{bagdasaryan2020backdoor}, the attacker makes duplicates of the local training examples on the malicious clients,  embeds the trigger to the duplicated training inputs, and assigns an attacker-chosen label to them. Then, model updates are computed based on the local training data augmented by such duplicated training examples. Furthermore, to amplify the impact of the model updates, the malicious clients further scale them up by a factor before reporting them to the server. In Distributed Backdoor Attack (DBA) \cite{xie2019dba}, the attacker decomposes the  trigger  into separate local patterns and embeds them into the local training data of different malicious clients. 
In A Little is Enough Attack \cite{baruch2019little}, the model updates on the malicious clients are first computed following the Scaling Attack \cite{bagdasaryan2020backdoor}. Then, the attacker crops the model updates to be in certain ranges so that the Byzantine-robust aggregation rules fail to eliminate their malicious effects.

\subsection{Byzantine-Robust FL Methods}

Roughly speaking, Byzantine-robust FL methods view clients' model updates as high dimensional vectors and apply robust methods to estimate the aggregated model update. 
Next, we review several popular Byzantine-robust FL methods. 

\myparatight{Krum \cite{blanchard2017machine}} 
Krum tries to find a single model update among the clients' model updates  as the aggregated model update in each iteration. The chosen model update is the one with the closest Euclidean distances to the nearest $n - k - 2$ model updates.

\myparatight{Trimmed-Mean and Median \cite{yin2018byzantine}}  Trimmed-Mean and Median are coordinate-wise aggregation rules that aggregate each coordinate of the model update separately. For each coordinate,  Trimmed-Mean first sorts the values of the corresponding coordinates in the clients' model updates. After removing the largest and the smallest $k$ values, Trimmed-Mean calculates the average of the remaining $n-2k$ values as the corresponding coordinate of the aggregated model update.  Median calculates the median value of the corresponding coordinates in all model updates and treats it as the corresponding coordinate of the aggregated model update.

\myparatight{FLTrust \cite{cao2020fltrust}} FLTrust  leverages an additional validation dataset on the server.  In particular, a local model update has a lower trust score if its update direction deviates more from that of the server model update calculated based on the validation dataset. However, it is nontrivial to collect a clean validation dataset and FLTrust has poor performance when the distribution of validation dataset diverges substantially from the overall training dataset.

\section{Problem Formulation}
We consider a typical FL setting in which $n$ clients collaboratively train a global model maintained on a cloud server. We suppose that each client has a local training dataset $D_i,~ i=1,2,\cdots,n$ and we use $D=\cup_{i=1}^nD_i$ to denote the joint training data. 
The optimal global model $w^*$ is a solution to the optimization problem: $w^*= {\rm arg~min}_w \sum_{i=1}^n f(D_i,w)$, where $f(D_i,w)$ is the loss for client $i$'s local training data. 
The FL process starts with an initialized global model $w_0$. At the beginning of each iteration $t$, the server first sends the current global model $w_t$ to the clients or a subset of them. A client $i$ then
computes the  gradient $g_i^t$ of its loss $f(D_i,w_t)$ with respect to $w_t$ and sends $g_i^t$ back to the server, where $g_i^t$ is the model update from client $i$ in the $t$th iteration. Formally, we have:
\begin{align}
\label{modelupdate}
    g_i^t = \nabla f(D_i,w_t).
\end{align}
We note that  client $i$ can also use stochastic gradient descent (SGD) instead of  gradient descent, perform SGD multiple steps locally, and send the accumulated gradients  back to the server as model update. However, we assume a client performs the standard gradient descent for one step for simplicity.  

After receiving the clients' model updates, the server computes a global model update $g^t$ via aggregating the  clients' model updates based on some aggregation rule. Then, the server updates the global model using the global model update,  i.e., $w_{t+1}=w_t- \alpha·g^t$, where $\alpha$ is the global learning rate. Different FL methods essentially use different aggregation rules. 

\myparatight{Attack model} We follow the attack settings in previous works \cite{fang2020local,cao2022mpaf,bagdasaryan2020backdoor,baruch2019little,xie2019dba}. Specifically, an attacker controls $m$ malicious clients, which can be fake clients injected by the attacker or genuine ones compromised by the attacker. However,
the server is not compromised. The attacker has the following background knowledge about the FL  system: local training data and model updates on the malicious clients, loss function, and learning rate. In each iteration $t$, each benign client  calculates and reports the true model update $g_i^t=\nabla f(D_i,w_t)$, while a malicious client sends carefully crafted  model update (i.e., $g_i^t\neq \nabla f(D_i,w_t)$) to the server. 

\myparatight{Problem definition} We aim to design a malicious-client detection method in the above FL setting. 
In each iteration $t$, the detection method takes clients'  model updates in the current and previous iterations as an input and classifies each client to be benign or malicious. When at least one client is classified as malicious by our method in a certain iteration, the server stops the FL process, removes the clients detected as malicious, and restarts the FL process on the remaining clients. Our goal is to detect majority of malicious clients as early as possible. 
After detecting and removing majority of malicious clients, Byzantine-robust FL methods can learn accurate global models since they are robust against the small number of malicious clients that miss detection. 

\section{FLDetector}

\subsection{Model-Updates Consistency}

A benign client $i$ calculates its model update $g_i^t$ in the $t$th iteration according to Equation~\ref{modelupdate}. Based on the Cauchy mean value theorem~\cite{lang1968second}, we have the following:
\begin{equation}
\label{hessian}
    g_i^t=g_i^{t-1} + \mathbf{H}_i^t\cdot(w_t - w_{t-1}), 
\end{equation}
where $\mathbf{H}_i^t = \int_0^1\mathbf{H}_i(w_{t-1}+x(w_t - w_{t-1}))dx$ is an integrated Hessian for client $i$ in iteration $t$,  $w_t$ is the global model in iteration $t$, and $w_{t-1}$ is the global model in iteration $t-1$.
Equation~\ref{hessian} encodes the consistency between client $i$'s model updates $g_i^t$ and $g_i^{t-1}$.
However, the integrated Hessian  $\mathbf{H}_i^t$ is hard to compute exactly. In our work, we use a L-BFGS algorithm \cite{ byrd1994representations} to approximate integrated Hessian. To be more efficient, we approximate a single integrated Hessian $\mathbf{\hat{H}}^t$ in each iteration $t$, which is used for all clients. 
Specifically, we denote by $\Delta w_t = w_t -w_{t-1}$ the global-model difference in iteration $t$, and we denote by $\Delta g^t = g_t - g_{t-1}$ the global-model-update difference in iteration $t$, where the global model update is  aggregated from the clients' model updates. We denote by $\Delta W_t = \{\Delta w_{t-N}, \Delta w_{t-N+1},\cdots,\Delta w_{t-1} \}$  the global-model differences in the past $N$ iterations, and we denote by $\Delta G_t = \{\Delta g_{t-N}, \Delta g_{t-N+1},\cdots,\Delta g_{t-1} \}$ the global-model-update differences in the past $N$ iterations in iteration $t$. 
Then, based on the L-BFGS algorithm, we can estimate $\mathbf{\hat{H}}^t$ using $\Delta W_t$ and $\Delta G_t$. For simplicity, we denote by $\mathbf{\hat{H}}^t = \text{L-BFGS} (\Delta W_t, \Delta G_t)$. Algorithm \ref{lbfgs} shows the specific implementation of L-BFGS algorithm in the experiments. The input to L-BFGS are $\textbf{v} = w_t-w_{t-1}$, $\Delta W_t$, and $\Delta G_t$. The output of L-BFGS algorithm is the projection of the Hessian matrix in the direction of $w_t-w_{t-1}$.

\begin{algorithm*}[t]
\caption{L-BFGS to Compute Hessian Vector Product}
\label{lbfgs}
\leftline{\textbf{Input}:  Global-model differences $\Delta W_t = \{\Delta w_{t-N}, \Delta w_{t-N+1},\cdots,\Delta w_{t-1} \}$, global-model-update  differences $\Delta G_t = \{\Delta g_{t-N},\Delta g_{t-N+1}, \cdots,$}
\leftline{$\Delta g_{t-1} \}$, vector $\mathbf{v} = w_t - w_{t-1}$, and window size $N$}
\leftline{\textbf{Output}:  Hessian vector product $\mathbf{\hat{H}}^t\mathbf{v}$}
\begin{algorithmic}[1] 
\STATE Compute $\Delta W_t^T \Delta W_t$\\ 
\STATE Compute $\Delta W_t^T \Delta G_t$, get its diagonal matrix $D_t$ and its lower triangular submatrix $L_t$\\
\STATE Compute $\sigma = \Delta g^T_{t-1} \Delta w_{t-1}/(\Delta w^T_{t-1} \Delta w_{t-1})$
\STATE Compute the Cholesky factorization for $\sigma \Delta W_t^T \Delta W_t + L_t D_t L_t^T $ to get $J_tJ_t^T$
\STATE Compute $q = \left[ \begin{array}{cc}
-D_t^{1/2} & D_t^{-1/2}L_t^T \\
0 & J_t^T \\
\end{array} 
\right ]^{-1} 
\left[ \begin{array}{cc}
D_t^{1/2} & 0 \\
D_t^{-1/2}L_t^T & J_t \\
\end{array} 
\right ]^{-1}
\left[ \begin{array}{c}
\Delta G_t^T \textbf{v}\\
\sigma \Delta W_t^T\textbf{v} \\
\end{array} 
\right ]$
\RETURN $\sigma \textbf{v} - \left[ \begin{array}{cc} \Delta G_t & \sigma \Delta W_t \end{array}\right]q$ 
\end{algorithmic}
\end{algorithm*}

Based on the estimated Hessian $\mathbf{\hat{H}}^t$, we \emph{predict} a client $i$'s model update in iteration $t$ as follows:
\begin{align}
\label{predictedmodelupdate}
    \hat{g}_i^t = g_i^{t-1} + \mathbf{\hat{H}}^t(w_t - w_{t-1}),
\end{align}
where $\hat{g}_i^t$ is the \emph{predicted model update} for client $i$ in iteration $t$. When the L-BFGS algorithm  estimates the integrated Hessian accurately, the predicted model update $\hat{g}_i^t$ is close to the actual model update ${g}_i^t$ for a \emph{benign} client $i$. In particular, if the estimated Hessian is exactly the same as the integrated Hessian, then the predicted model update equals the actual model update  for a benign client. However, no matter whether the integrated Hessian is estimated accurately or not, the predicted model update would be different from the model update sent by a malicious client. In other words, the predicted model update and the received one are consistent for benign clients but inconsistent for malicious clients, which we leverage to detect malicious clients.

\begin{algorithm}[t]
\caption{Gap Statistics}
\label{gap statistics}
\leftline{\textbf{Input}: Clients' suspicious scores $s^t$, number of sampling $B$,} 
\leftline{maximum number of clusters $K$, and number of clients $n$.}
\leftline{\textbf{Output}: Number of clusters $k$.}
\begin{algorithmic}[t] 
\FOR{$k$ = 1,2, $\cdots$, $K$}
    \STATE Apply linear transformation on $s^t$ so that the minimum of $s^t$ equals 0 and the maximum of $s^t$ equals 1.
    \STATE Apply $k$-means  on the suspicious scores to get clusters $\{C_i\}$ and means $\{\mu_i\}$.
    \STATE $V_k = \sum_{i=1}^k \sum_{x_j \in C_i} \|x_j-\mu_i\|^2$
    \FOR{$b =$ 1,2, $\cdots$, $B$}
      \STATE Sample $n$ points uniformly in [0,1] 
         \STATE Perform $k$-means and calculate 
         \STATE $V^*_{kb}= \sum_{i=1}^k \sum_{x_{jb} \in C_{ib}} \|x_{jb}-\mu_{ib}\|^2$
    \ENDFOR
    \STATE $Gap(k) = \frac{1}{B}\sum_{i=1}^B log(V^*_{kb})- log (V_k)$
    \STATE $v'=\frac{1}{B}\sum_{i=1}^B log(V^*_{kb})$
    \STATE $sd(k) = (\frac{1}{B}\sum_{i=1}^B (log(V^*_{kb})) - v')^2)^{\frac{1}{2}}$
    \STATE $s'_k = \sqrt{\frac{1+B}{B}}sd(k)$
\ENDFOR
\STATE $\hat{k}$ = smallest $k$ such that $Gap(k) - Gap(k+1)+s'_{k+1} \ge 0.$ 
\RETURN $\hat{k}$ 
\end{algorithmic}
\end{algorithm}

\subsection{Detecting Malicious Clients}

\textbf{Suspicious score for a client:} Based on the model-updates consistency discussed above, we assign a suspicious score for each client. Specifically, we measure the consistency between a predicted model update $\hat{g}_i^t$ and a received model update ${g}_i^t$ using their Euclidean distance. We denote by $d^t$ the vector of such $n$ Euclidean distances for the $n$ clients in iteration $t$, i.e., $d^t=[\| \hat{g}_1^t-{g}_1^t\|_2, \| \hat{g}_2^t-{g}_2^t\|_2, \cdots, \| \hat{g}_n^t-{g}_n^t\|_2]$. We normalize the vector $d^t$ as $\hat{d}^t=d^t/\|d^t\|_1$. We use such normalization to incorporate the model-updates consistency variations across different iterations.  
Finally, our suspicious score $s_i^t$ for client $i$ in iteration $t$ is the client's average  normalized Euclidean distance in the past $N$ iterations, i.e., $s_i^t=\frac{1}{N}\sum_{r=0}^{N-1} \hat{d}_i^{t-r}$. We call $N$ \emph{window size}. 

\myparatight{Unsupervised detection via $k$-means} In iteration $t$, we perform malicious-clients detection based on the clients' suspicious scores $s_1^t, s_2^t, \cdots, s_n^t$. Specifically, we cluster the clients based on their suspicious scores  $s_1^t, s_2^t, \cdots, s_n^t$, and we use the Gap statistics \cite{tibshirani2001estimating} to determine the number of clusters. If the clients can be grouped into more than 1 cluster based on the Gap statistics, then we use $k$-means to divide the clients into 2 clusters based on their suspicious scores. Finally, the clients in the cluster with larger average suspicious score are classified as  malicious. When at least one client is classified as malicious in a certain iteration, the detection  finishes, and the server removes the clients classified as malicious and restarts the training.

Algorithm \ref{gap statistics} shows the pseudo codes of Gap statistics algorithm. The input to Gap statistics are the vectors of suspicious scores $s^t$, the number of sampling $B$, the number of maximum clusters $K$, and the number of clients $n$. The output of Gap statistics is the number of clusters $K$. Generally, Gap statistics compares the change in within-cluster dispersion with that expected under a reference null distribution, i.e., uniform distribution, to determine the number of clusters. The computation of the gap statistic involves the following steps: 1) Vary the number of clusters $k$ from 1 to $K$ and cluster the suspicious scores with $k$-means. Calculate $W_k = \sum_{i=1}^k \sum_{x_j \in C_i} \|x_j-\mu_i\|^2$. 2) Generate $B$ reference data sets and cluster each of them with $k$-means. Compute the estimated gap statistics $Gap(k) = \frac{1}{B}\sum_{i=1}^B log(W^*_{kb})- log (W_k)$. 3) Compute the standard deviation $sd(k) = (\frac{1}{B}\sum_{i=1}^B (log(W^*_{kb})) - w')^2)^{\frac{1}{2}}$ and define $s_{k+1} = \sqrt{\frac{1+B}{B}}sd(k)$. 4) Choose the number of clusters $\hat{k}$ as the smallest $k$ such that $Gap(k) - Gap(k+1)+s_{k+1} \ge 0.$ If there are more than one cluster, the attack detection $Flag$ is set to positive because there are outliers in the suspicious scores.

Algorithm~\ref{pseudo code} summarizes the algorithm of FLDetector.
\begin{algorithm}[t]
\caption{FLDetector}
\label{pseudo code}
\leftline{\textbf{Input}: Total training iterations $Iter$ and window size $N$.}
\leftline{\textbf{Output}: Detected malicious clients or none.}

\begin{algorithmic}[1] 
\FOR{$t = 1, 2, \cdots, Iter$}
    \STATE $\mathbf{\hat{H}}^t = \text{L-BFGS} (\Delta W_t, \Delta G_t)$. \\
    \FOR{$i = 1, 2, \cdots, n$}
    \STATE $\hat{g}_i^t = g_i^{t-1} + \mathbf{\hat{H}}^t(w_t - w_{t-1})$. \\
    \ENDFOR
     \STATE $d^t=[\| \hat{g}_1^t-{g}_1^t\|_2, \| \hat{g}_2^t-{g}_2^t\|_2, \cdots, \| \hat{g}_n^t-{g}_n^t\|_2]$. \\
    \STATE  $\hat{d}^t=d^t/\|d^t\|_1$.
    \STATE $s_i^t=\frac{1}{N}\sum_{r=0}^{N-1} \hat{d}_i^{t-r}$. 
    \STATE Determine the number of clusters $k$ by Gap statistics.

\IF   {   $k > 1 $  }  
\STATE   Perform $k$-means clustering based on the suspicious scores with $k = 2$. \\  
\RETURN    The clients in the cluster with larger average suspicious score as malicious.  
\ENDIF
\ENDFOR
\RETURN  None. 
\end{algorithmic}
\end{algorithm}

\subsection{Complexity Analysis}
To compute the estimated Hessian, the server needs to save the global-model differences and global-model-update differences in the latest $N$ iterations. Therefore, the storage overhead of FLDetector for the server is $O(Np)$, where $p$ is the number of parameters in the global model. Moreover, according to \cite{byrd1994representations}, the complexity of estimating the Hessian $\mathbf{\hat{H}}^t$ using L-BFGS and computing the Hessian vector product $\mathbf{\hat{H}}^t(w_t - w_{t-1})$ is $O(N^3 + 6Np)$ in each iteration. The complexity of calculating the suspicious scores is $O(2np+ Nn)$ in each iteration, where $n$ is the number of clients. The total complexity of Gap statistics and $k$-means is $O(KBn^2)$ where $K$ and $B$ are the number of maximum clusters and sampling in Gap statistics. Therefore, the total time complexity of FLDetector in each iteration is $O(N^3 + KBn^2 +(6N+ 2n)p +Nn)$. Typically, $K$, $B$, $n$, and $N$ are much smaller than $p$. Thus, the time complexity of FLDetector for the server is roughly  linear to the number of  parameters in the global model in each iteration. We note that the server is powerful in FL, so the storage and computation overhead of FLDetector for the server is acceptable. As for the clients, FLDetector does not incur extra computation and communication overhead.

\subsection{Theoretical Analysis on Suspicious Scores}
We compare the suspicious scores of benign and malicious clients theoretically. We first describe the definition of $L$-smooth gradient, which is widely used for theoretical analysis on  machine learning. 

\begin{definition}
We say a client's loss function is $L$-smooth if  we have the following inequality for any $~\bm{w}$ and $\bm{w}'$:
\begin{equation}
 \Vert \nabla f(D_i,\bm{w})-\nabla f(D_i,\bm{w}')\Vert \leq L\Vert \bm{w}-\bm{w}'\Vert,   
\end{equation}
where $f(D_i,w)$ is the client's loss function  and $\Vert\cdot \Vert$ represents $\ell_2$ norm of a vector. 
\label{as:convex}
\end{definition}

\begin{thm}
Suppose the gradient of each client's loss function is $L$-smooth, 
FedAvg is used as the aggregation rule, the clients' local training datasets are iid,  the learning rate $\alpha$ satisfies $\alpha \textless \frac{1}{(N+2)L}~$($N$ is the window size). Suppose the malicious clients perform an untargeted model poisoning attack in each iteration by reversing the true model updates as the poisoning ones, i.e., each malicious client $i$ sends $-g_i^t$ to the server in each iteration $t$.  Then we have the expected suspicious score of a benign client is smaller than that of a malicious client in each iteration $t$. Formally, we have the following inequality: 
\begin{equation}
    \mathbb{E}(s_i^t) < \mathbb{E}(s_j^t), \forall i\in \mathcal{B}, \forall j \in \mathcal{M}, 
\end{equation}
where the expectation $\mathbb{E}$ is taken with respect to the randomness in the clients' local training data, $\mathcal{B}$ is the set of benign clients, and $\mathcal{M}$ is the set of malicious clients.
\label{thm}
\end{thm}
\begin{proof}
Our idea is to bound the difference between predicted model updates and the received ones from benign clients. Appendix shows our detailed proof. 
\end{proof}

\subsection{Adaptive Attacks}
When the attacker knows that our FLDetector is used to detect malicious clients, the attacker can adapt its attack to FLDetector to evade detection. Therefore, we design and evaluate adaptive attacks to FLDetector. Specifically, we formulate an adaptive attack by adding an extra term to regularize the loss function used to perform existing attacks. Our regularization term measures the Euclidean distance between a predicted model update and a local model update. Formally, a malicious client $i$ solves the following optimization problem to perform an adaptive attack in iteration $t$:
\begin{equation}
    \mathop{\rm min}\limits_{g_i^t} \lambda \mathcal{L}_{attack}+ (1-\lambda)\Vert{g}_i^t- ({g}_i^{t-1} + \mathbf{\hat{H}}^t_i (w_t - w_{t-1}))\Vert,
\end{equation}
where $\mathcal{L}_{attack}$ is the loss function used to perform existing attacks~\cite{fang2020local, xie2019dba, baruch2019little, bagdasaryan2020backdoor}, ${g}_i^t$ is the poisoning local model update on malicious client $i$ in iteration $t$, ${g}_i^{t-1} + \mathbf{\hat{H}}^t_i (w_t - w_{t-1})$ is the predicted model update for client $i$, and $\mathbf{\hat{H}}^t_i$ is the Hessian calculated on client $i$'s dataset to approximate $\mathbf{\hat{H}}^t$. $\lambda \in (0, 1]$ is a hyperparameter to balance the loss function and the regularization term. 
A smaller $\lambda$ makes the malicious clients less likely to be detected, but the attack is also less effective. 

\section{Experiments}

\subsection{Experimental Setup}
\textbf{Datasets and global-model architectures:}
We consider three widely-used benchmark datasets {MNIST} \cite{mnist}, {CIFAR10} \cite{cifar}, and {FEMNIST} \cite{caldas2018leaf} to evaluate FLDetector. For MNIST and CIFAR10, we assume there are 100 clients and use the method in \cite{fang2020local} to distribute the training images to the clients. Specifically, this method has a parameter called \emph{degree of non-iid} ranging from 0.1 to 1.0 to control the distribution of the clients' local training data. The clients’ local training data are not \emph{independent and identically distributed (iid)} when the degree
of non-iid is larger than 0.1 and are more non-iid when the  degree
of non-iid becomes larger. Unless otherwise mentioned, we set the degree of non-iid  to 0.5. FEMNIST is a 62-class classification dataset from the open-source benchmark library of FL \cite{caldas2018leaf}. The training images are already grouped by the writers and we randomly sample 300 writers, each of which is treated as a client. We use a four-layer Convolutional Neural Network (CNN) (see Table~\ref{cnn})  as the global model  for MNIST and FEMNIST. For  CIFAR-10, we consider the widely used ResNet20 architecture \cite{he2016deep} as the global model.

\begin{table}[!t]
\centering
\caption{The CNN architecture of the global model used for MNIST and FEMNIST.}
\begin{tabular}{|c|c|}

\hline
Layer                  & Size                 \\ \hline
Input                  & 28$\times$ 28 $\times$ 1 \\ \hline
Convolution + ReLU     & 3$\times$ 3 $\times$ 30  \\ \hline
Max Pooling            & 2 $\times$ 2           \\ \hline
Convolution + ReLU     & 3 $\times$ 3 $\times$ 5  \\ \hline
Max Pooling            & 2 $\times$ 2           \\ \hline
Fully Connected + ReLU & 100                  \\ \hline
Softmax                & 10 (62 for FEMNIST)  \\ \hline
\end{tabular}
\label{cnn}
\end{table}

\myparatight{FL settings}
 We consider four FL methods: FedAvg \cite{mcmahan2017communication}, Krum \cite{blanchard2017machine}, Trimmed-Mean \cite{yin2018byzantine}, and Median \cite{yin2018byzantine}. We didn't consider FLTrust \cite{cao2020fltrust}  due to its additional requirement of a clean validation dataset. 
 Considering the different characteristics of the datasets, we adopt the following parameter settings for  FL training: for MNIST, we train 1,000 iterations with a learning rate of $2\times {10}^{-4}$; and for CIFAR10 and FEMNIST, we train 2,000 iterations with a learning rate of $1\times{10}^{-3}$. For simplicity, we assume all clients are involved in each iteration of FL training. 
Note that when FLDetector detects  malicious clients in a certain iteration, the server removes the clients classified as malicious, restarts the FL training, and repeats for the pre-defined number of iterations. 

\begin{table*}[t]
\centering
\caption{DACC, FPR, and FNR of malicious-client detection for different attacks, detection methods, and aggregation rules. The best detection results are bold for each attack. FEMNIST dataset, CNN global model, and 28 malicious clients are used.}

\small
\begin{tabular}{ccp{0.42cm}<{\centering}p{0.42cm}<{\centering}p{0.42cm}<{\centering}p{0.42cm}<{\centering}p{0.42cm}<{\centering}p{0.42cm}<{\centering}p{0.42cm}<{\centering}p{0.42cm}<{\centering}p{0.42cm}<{\centering}p{0.42cm}<{\centering}p{0.42cm}<{\centering}p{0.42cm}<{\centering}}
\toprule
\multirow{2}{*}{Attack}                                                                     & \multirow{2}{*}{Detector}                                     & \multicolumn{3}{c}{FedAvg} &  \multicolumn{3}{c}{Krum}& \multicolumn{3}{c}{Trimmed-Mean} & \multicolumn{3}{c}{Median}  \\ \cmidrule(lr){3-5}\cmidrule(lr){6-8}\cmidrule(lr){9-11}\cmidrule(lr){12-14}
                                                                                            &                                                              & \scriptsize DACC     &  FPR     &  FNR    & \scriptsize DACC      & FPR      & FNR     & \scriptsize DACC     & FPR     & FNR    & \scriptsize DACC    & FPR    & FNR    \\ \midrule
\multirow{4}{*}{\begin{tabular}[c]{@{}c@{}}Untargeted\\ Model\\ Poisoning\\ Attack\end{tabular}} 
& VAE                               &0.71 & 0.02 & 0.99 & 0.57 & 0.36 & 0.62 & 0.56 & 0.37 & 0.62 & 0.55 & 0.35 & 0.71   \\
& FLD-Norm                                                        & 0.72 & 0.03 & 0.93 & 0.05 & 0.93 & 1.00 & 0.42 & 0.42 & 1.00 & 0.13 & 0.82 & 1.00    \\
                                                                                            & FLD-NoHVP& 0.51 & 0.38 & 0.79 & 0.34 & 0.83 & 0.21 & 0.77 & 0.32 & 0.00 & 0.67 & 0.28 & 0.54    \\
                                                                                            & FLDetector                                                 & \textbf{1.00}       & \textbf{0.00}       & \textbf{0.00}       & \textbf{1.00}       & \textbf{0.00}       & \textbf{0.00}        & \textbf{1.00}       & \textbf{0.00}       & \textbf{0.00}       & \textbf{1.00}       & \textbf{0.00}       & \textbf{0.00}             \\ \midrule
\multirow{4}{*}{\begin{tabular}[c]{@{}c@{}}Scaling\\ Attack\end{tabular}}                                                                          & VAE                                                          &0.73 & 0.05 & 0.99 & 0.68 & 0.44 & 0.00 & 0.33 & 0.54 & 1.00 & 0.47 & 0.42 & 0.82   \\
& FLD-Norm                                                         & 0.82 & 0.14 & 0.29 & 0.68 & 0.44 & 0.00 & 0.92 & 0.00 & 0.29 & 0.90 & 0.03 & 0.29     \\
                                                                                            & FLD-NoHVP & 0.07 & 0.98 & 0.82 & 0.42 & 0.42 & 1.00 & 0.91 & 0.13 & 0.00 & 0.96 & 0.05 & 0.00  \\
                                                                                            & FLDetector                                                   & \textbf{0.85}       & \textbf{0.20}       & \textbf{0.00}       & \textbf{1.00}        & \textbf{0.00}   & \textbf{0.00}        & \textbf{0.98}       & \textbf{0.03}       & \textbf{0.00}       & \textbf{1.00}    & \textbf{0.00 }     & \textbf{0.00}   \\ \midrule
\multirow{4}{*}{\begin{tabular}[c]{@{}c@{}}Distributed\\Backdoor\\Attack\end{tabular}}                                             & VAE                 & 0.75 & 0.07 & 0.71 & 0.69 & 0.43 & 0.00 & 0.52 & 0.28 & 1.00 & 0.53 & 0.68 & 1.00        \\
& FLD-Norm         & 0.66 & 0.33 & 0.36 & 0.65 & 0.42 & 0.18 & 0.73 & 0.28 & 0.25 & 0.75 & 0.22 & 0.33 \\
                                                                                            & FLD-NoHVP & 0.09 & 0.98 & 0.75 & 0.46 & 0.64 & 0.29 & 0.90 & 0.11 & 0.07 & 0.98 & 0.03 & 0.00           \\
                                                                                            & FLDetector                                      & \textbf{0.92}       & \textbf{0.11}       & \textbf{0.00}       & \textbf{1.00}        & \textbf{0.00}        & \textbf{0.00}        & \textbf{1.00}       & \textbf{0.00}       & \textbf{0.00}       & \textbf{1.00}    & \textbf{0.00}      & \textbf{0.00 }       \\ \midrule
\multirow{4}{*}{\begin{tabular}[c]{@{}c@{}}A Little\\is Enough\\Attack\end{tabular}}         
&VAE                                                        & 0.80 & 0.22 & 0.14 & 0.77 & 0.71 & 0.11 & 0.92 & 0.00 & 0.29 & 0.93 & 0.00 & 0.25   \\& FLD-Norm  & 0.05 & 0.93 & 1.00 & 0.11 & 0.97 & 0.68 & 0.02 & 0.97 & 1.00 & 0.08 & 0.89 & 1.00       \\ 
                                                                                            & FLD-NoHVP & 0.49 & 0.40 & 0.79 & 0.47 & 0.35 & 1.00 & 0.23 & 0.69 & 0.96 & 0.26 & 0.69 & 0.86  \\
& FLDetector                                                   & \textbf{0.93}       & \textbf{0.10}       & \textbf{0.00}       & \textbf{1.00}        & \textbf{0.00}        & \textbf{0.00}        & \textbf{1.00}       & \textbf{0.00}       & \textbf{0.00}       & \textbf{1.00}    & \textbf{0.00}      & \textbf{0.00}   \\ \bottomrule
\end{tabular}
\label{detection2}
\end{table*}

\begin{table*}[t]
\small
\centering
\caption{TACC and ASR of the global models learnt by Median in different scenarios. The results for the targeted model poisoning attacks are in the form of “TACC / ASR ($\%$)”. 28 malicious clients are used.} 

\begin{tabular}{llccc}
\toprule
Dataset                 & Attack                & No Attack &  \begin{tabular}[c]{@{}c@{}} w/o FLDetector \\ \end{tabular} & \begin{tabular}[c]{@{}c@{}} w/ FLDetector\end{tabular}\\ \midrule
\multirow{4}{*}{MNIST}   &Untargeted Model Poisoning Attack   &97.6    & 69.5                  & 97.4                 \\
                         & Scaling Attack                &97.6     & 97.6/0.5              & 97.6/0.5             \\
                         & Distributed Backdoor Attack   &97.6     & 97.4/0.5              & 97.5/0.4             \\
                         & A Little is Enough Attack        &97.6    & 97.8/100.0            & 97.9/0.3             \\ \midrule
\multirow{4}{*}{CIFAR10} &Untargeted Model Poisoning Attack   &65.8    & 27.8                  & 65.9                 \\
                         & Scaling Attack                &65.8    & 66.6/91.2             & 65.7/2.4             \\
                         & Distributed Backdoor Attack    &65.8    & 66.1/93.5             & 65.2/1.9             \\
                         & A Little is Enough Attack           &65.8    & 62.1/95.2             & 64.3/1.8             \\ \midrule
\multirow{4}{*}{FEMNIST} &Untargeted Model Poisoning Attack   &64.4    & 14.3                  & 63.2                 \\
                         & Scaling Attack                &64.4    & 66.4/57.9             & 64.5/1.7             \\
                         & Distributed Backdoor Attack    &64.4    & 67.5/53.2             & 64.3/2.1             \\
                         & A Little is Enough Attack           &64.4    & 66.7/59.6             & 65.0/1.6             \\ \bottomrule
\end{tabular}
\label{end-to-end}
\end{table*}

\myparatight{Attack settings}
By default, we randomly sample 28$\%$ of the
clients as malicious ones. We choose this fraction because in the Distributed Backdoor Attack (DBA), the trigger pattern need to be equally splitted into four parts and embedded into the local training data of four malicious clients groups. Specifically, the number of malicious
clients is 28, 28, and 84 for MNIST, CIFAR10, and FEMNIST, respectively. We consider one Untargeted Model Poisoning Attack \cite{fang2020local}, as well as three targeted model poisoning attacks including Scaling Attack \cite{bagdasaryan2020backdoor}, Distributed Backdoor Attack \cite{xie2019dba}, and A Little is Enough Attack \cite{baruch2019little}. 
For all the three targeted model poisoning attacks, the trigger patterns are the same as  their original papers and label '0' is selected as the target label. 
The scaling factor is set to 100 following \cite{bagdasaryan2020backdoor}.
Unless otherwise mentioned, the malicious clients perform attacks in \emph{every iteration} of FL training.\par

\myparatight{Compared detection methods}  There are few works on detecting malicious clients in FL. We compare the following methods:
\begin{itemize}
    \item {\bf VAE \cite{li2020learning}.} This method  trains a variational autoencoder for benign model updates by simulating model training  using a validation dataset on the server and then applies it to detect malicious clients during FL training. We consider the validation dataset is the same as the joint local training data of all clients, which gives a strong advantage to VAE. 
    
    \item {\bf FLD-Norm.} This is a variant of FLDetector. Specifically, FLDetector considers the Euclidean distance between a predicted model update and the received one in suspicious scores. One natural question is whether the norm of a model update itself can be used to detect malicious clients. In FLD-Norm, the distance vector $d^t$ consists of the $\ell_2$ norms of the $n$ clients' model updates in iteration $t$, which are further normalized and used to calculate our suspicious scores. 
    
    \item {\bf FLD-NoHVP.} This is also a variant of FLDetector. In particular, in this variant, we do not consider the Hessian vector product (HVP) term in Equation~\ref{predictedmodelupdate}, i.e., $\hat{g}_i^t = g_i^{t-1}$. The clients' suspicious scores are calculated based on such predicted model updates.  We use this variant to show that the Hessian vector product term in predicting the model update is important for FLDetector. 
    
 \end{itemize}

\myparatight{Evaluation metrics} We consider evaluation metrics for both detection and the learnt global models. For detection, we  use \emph{detection accuracy (DACC)}, \emph{false positive rate (FPR)}, and \emph{false negative rate (FNR)} as evaluation metrics. DACC is the fraction of clients that are correctly classified as benign or malicious. 
FPR (or FNR) is the fraction of benign (or malicious) clients that are falsely classified as malicious (or benign).  
To evaluate the learnt global model, we use \emph{testing accuracy (TACC)}, which is the fraction of testing examples that are correctly classified by the global model. Moreover, for targeted model poisoning attacks, we further use \emph{attack success rate (ASR)} to evaluate the global model. In particular, we embed the trigger to each testing input and the ASR is the fraction of trigger-embedded testing inputs that are classified as the target label by the global model. A lower ASR means that a  targeted model poisoning attack is less successful.

\myparatight{Detection settings} By default, we start to detect malicious clients in the 50th iteration of FL training, as we found the first dozens of iterations may be unstable. We will show how the iteration to start detection affects the performance of FLDetector. If no malicious clients are detected after finishing training for the pre-defined number of iterations, we classify all clients as benign. We set the window size $N$  to 10. Moreover, we set the maximum number of clusters $K$ and number of sampling $B$ in Gap statistics to 10 and 20, respectively. We will also explore the impact of hyperparameters in the following section.

\subsection{Experimental Results}
\textbf{Detection results:}  Table \ref{detection2} shows  the detection results  on the FEMNIST dataset for different attacks, detection methods, and FL methods. The results on  MNIST and CIFAR10  are respectively shown in Table~\ref{detection} and Table~\ref{detection1} in the Appendix,  due to limited space. 
We have several observations. First, FLDetector can detect majority of the malicious clients. For instance, on FEMNIST, the FNR of FLDetector is always 0.0  for different attacks and FL methods. Second,  FLDetector falsely detects a small fraction of benign clients as malicious, e.g., the  FPR of FLDetector ranges between 0.0 and 0.20 on FEMNIST for different attacks and FL methods. Third, on FEMNIST, FLDetector outperforms VAE for different attacks and FL methods; on MNIST and CIFAR10, FLDetector outperforms VAE in most cases and achieves comparable performance in the remaining cases. 
Fourth, FLDetector outperforms the two variants in most cases while achieving comparable performance in the remaining cases, which means that model-updates consistency and the Hessian vector product in estimating the model-updates consistency are informative at detecting malicious clients. Fifth, FLDetector achieves higher DACC for Byzantine-robust FL methods (Krum, Trimmed-Mean, and Median) than for FedAvg. The reason may be  that Byzantine-robust FL methods provide more robust  estimations of global model updates under attacks, which makes the estimation of Hessian and FLDetector more accurate.  

\begin{figure*}[t]
	\centering
	\subfigure[Untargeted Model Poisoning Attack]{\includegraphics[width=0.24\linewidth]{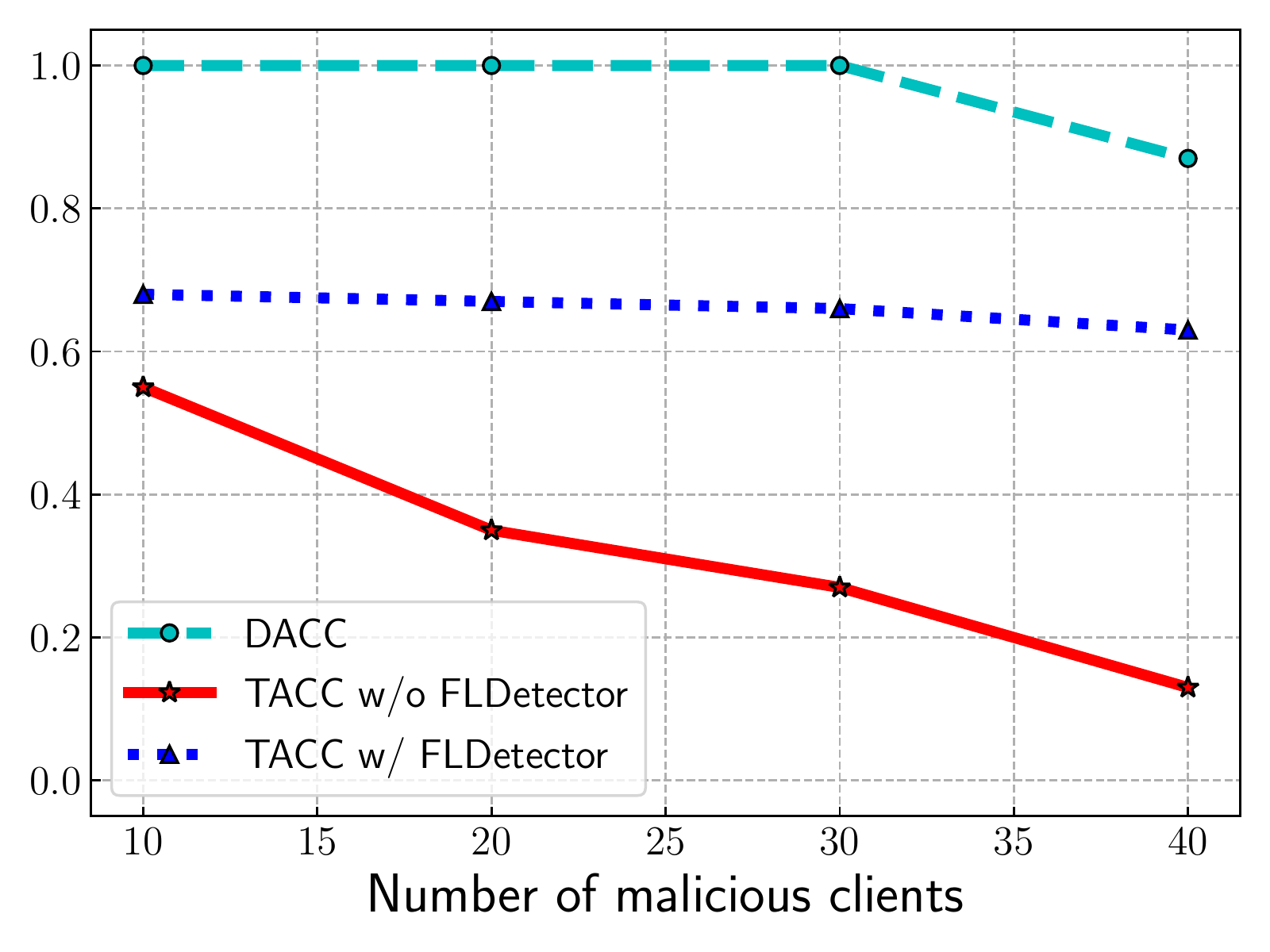}}
    \subfigure[Scaling Attack]{\includegraphics[width=0.24\linewidth]{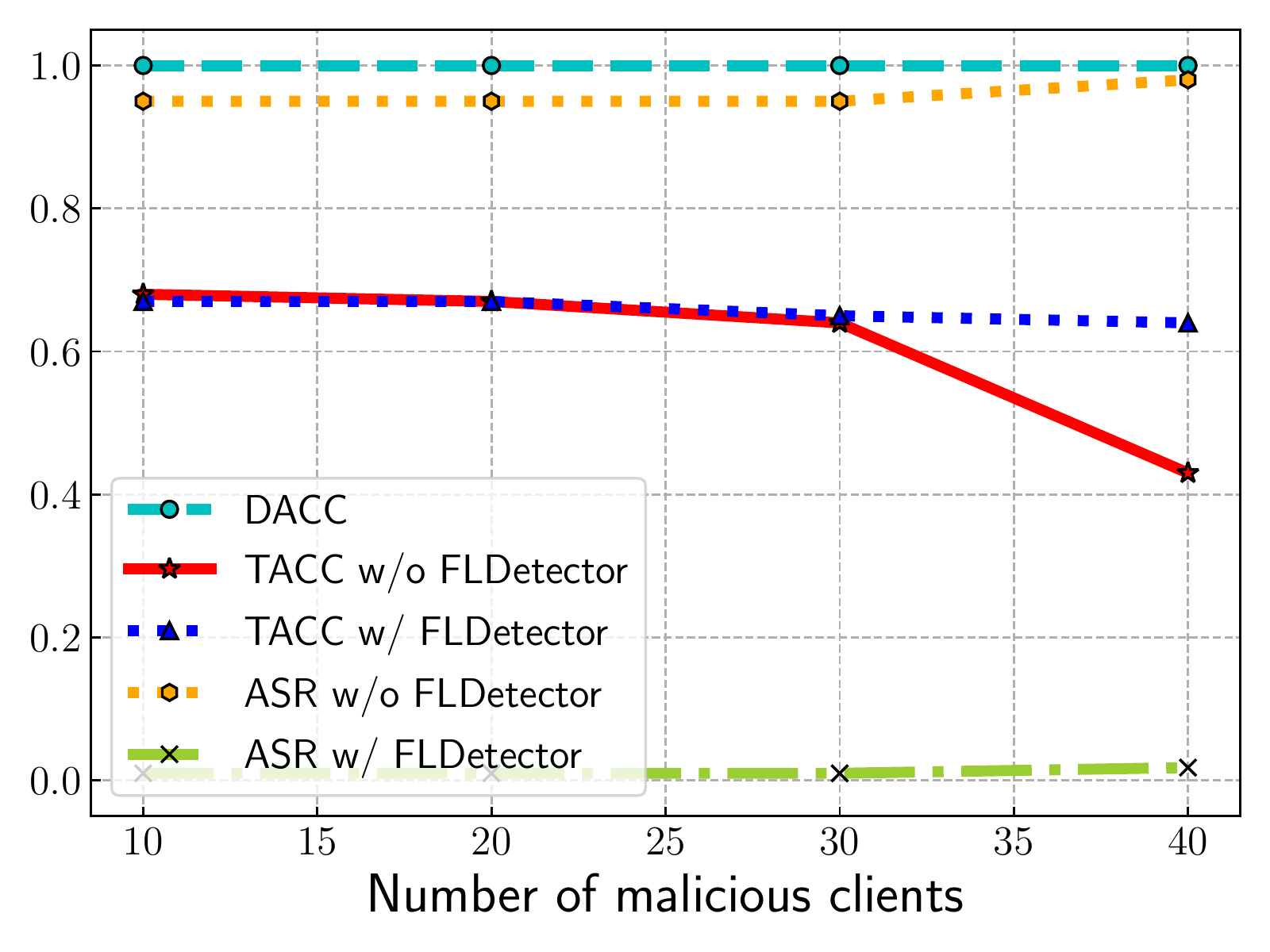}}
    \subfigure[Distributed Backdoor Attack]{\includegraphics[width=0.24\linewidth]{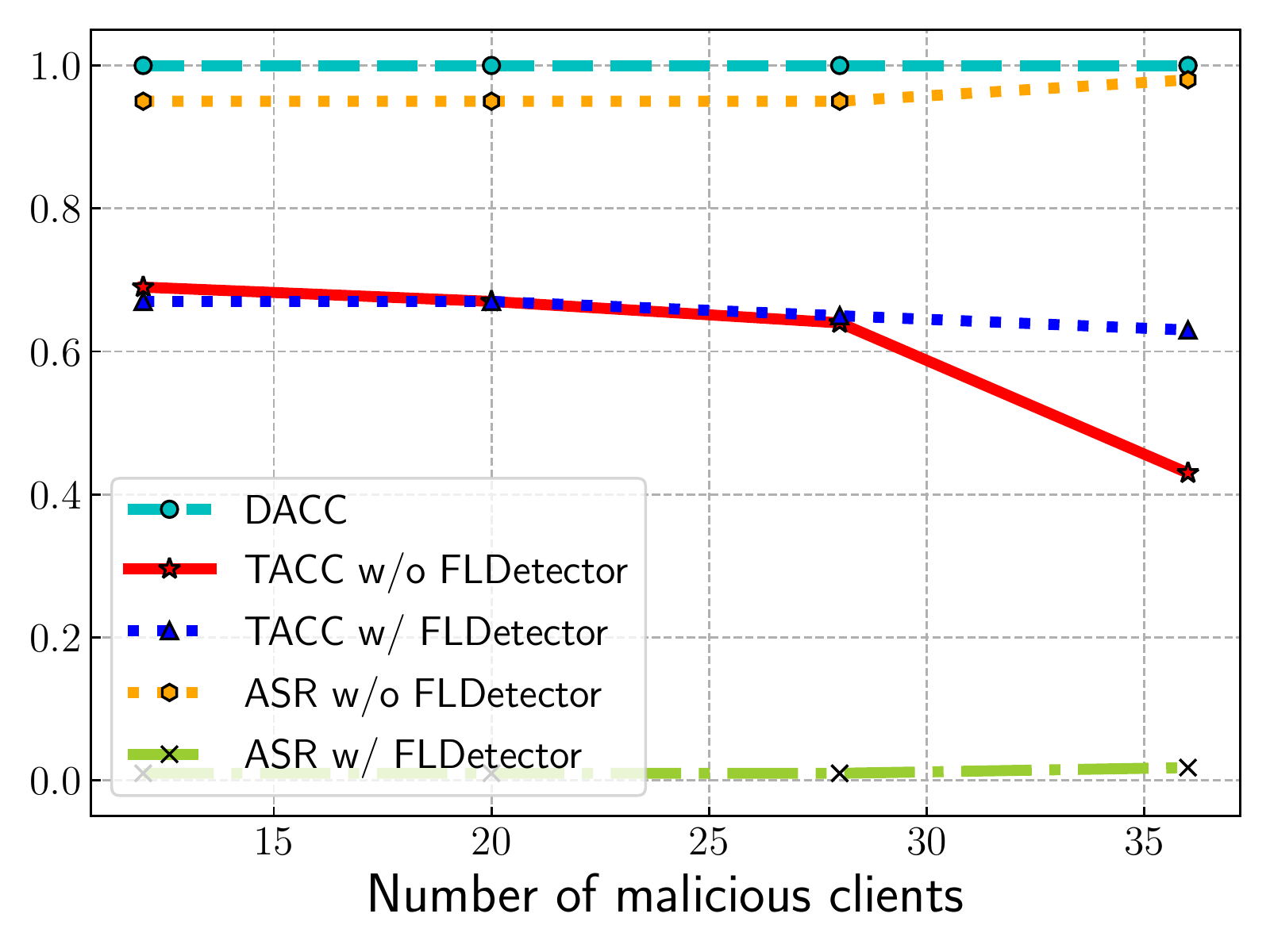}}
    \subfigure[A Little is Enough Attack]{\includegraphics[width=0.24\linewidth]{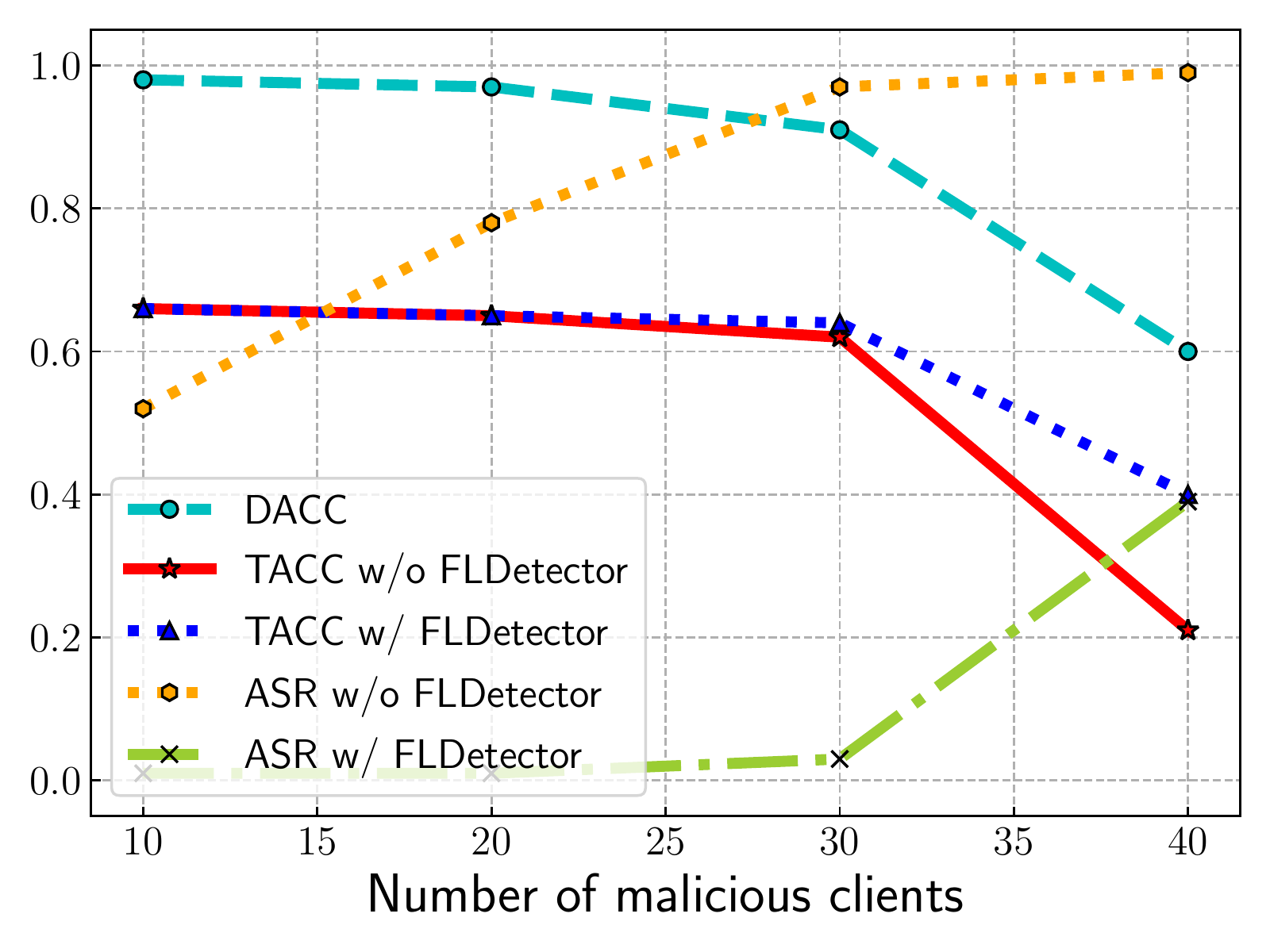}}
	    \vspace{-2mm}
	\caption{Impact of the number of malicious clients on FLDetector, where CIFAR10, Median, and 0.5 degree of non-iid are used.}
	\label{client}
\end{figure*}

\begin{figure*}[t]
	\centering
	\subfigure[Untargeted Model Poisoning Attack]{\includegraphics[width=0.24\linewidth]{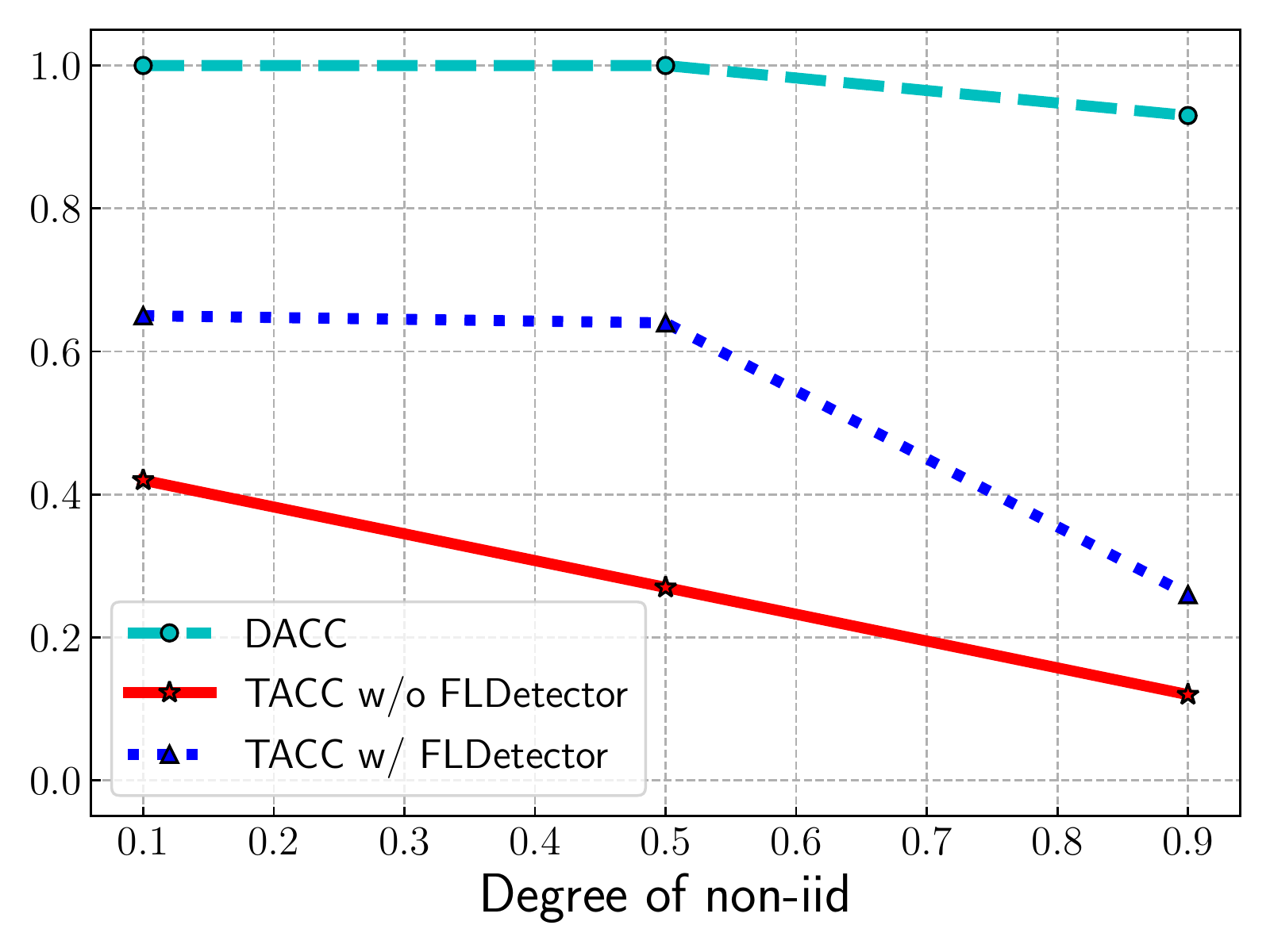}}
    \subfigure[Scaling Attack]{\includegraphics[width=0.24\linewidth]{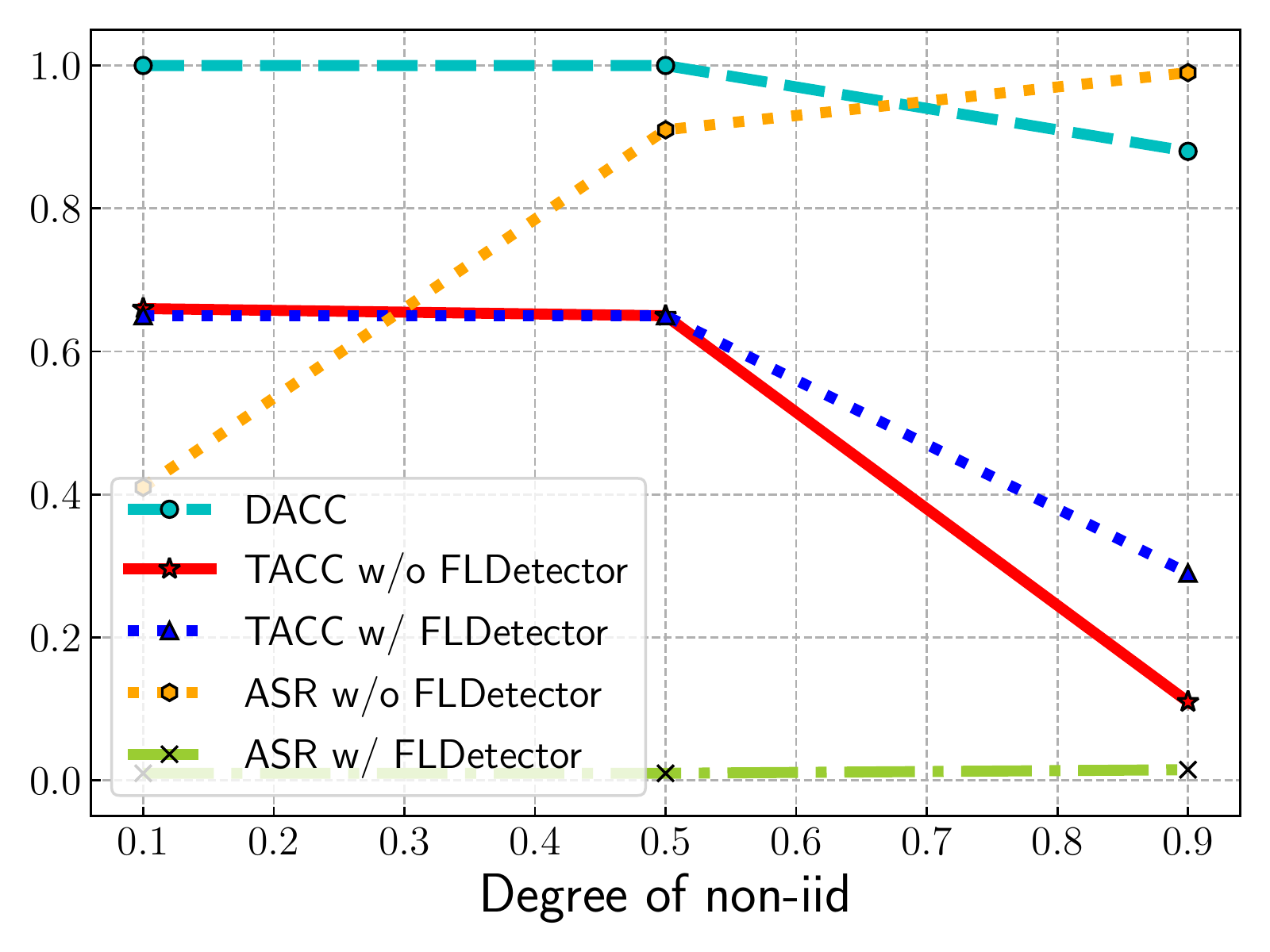}}
    \subfigure[Distributed Backdoor Attack]{\includegraphics[width=0.24\linewidth]{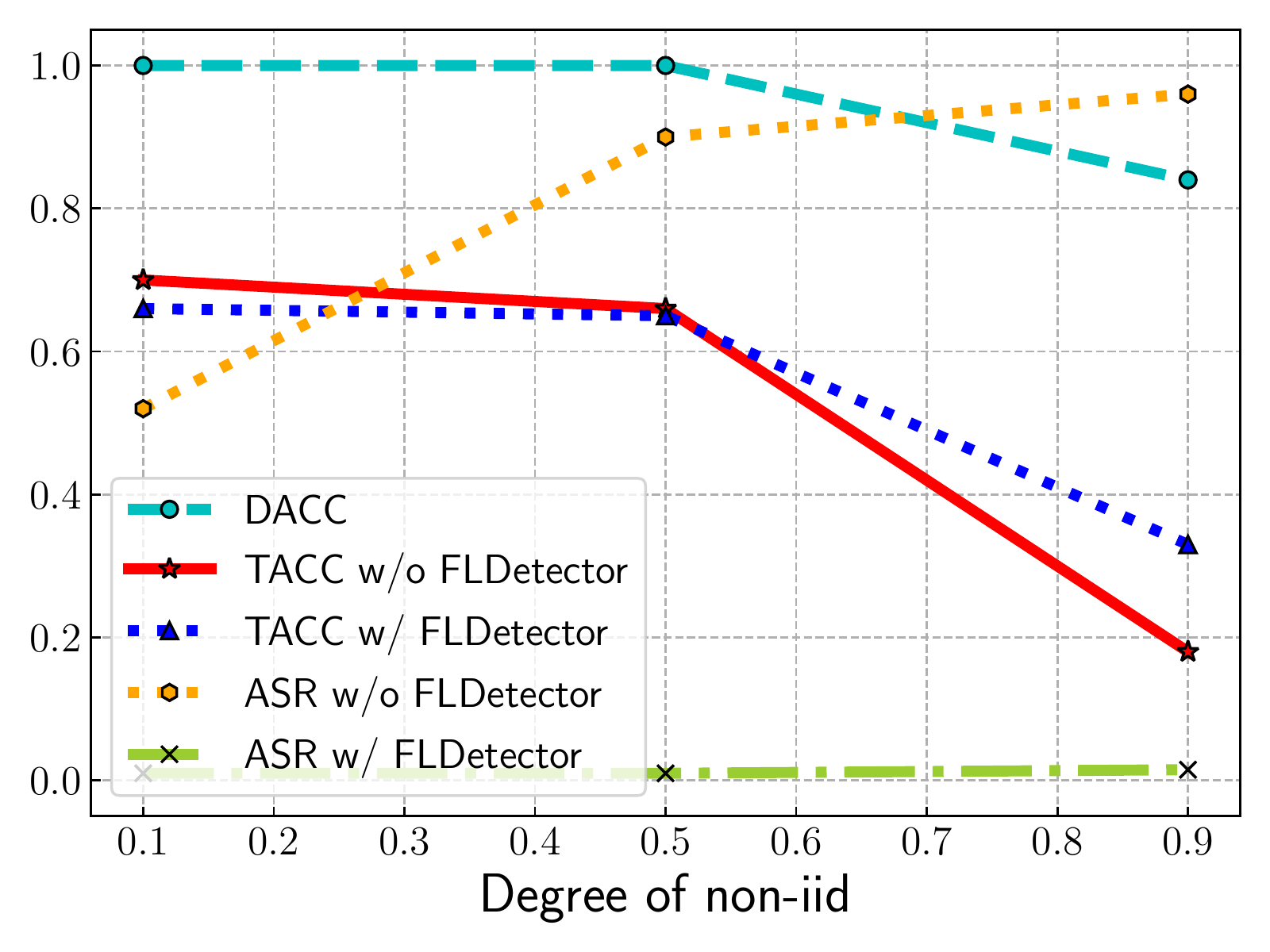}}
    \subfigure[A Little is Enough Attack]{\includegraphics[width=0.24\linewidth]{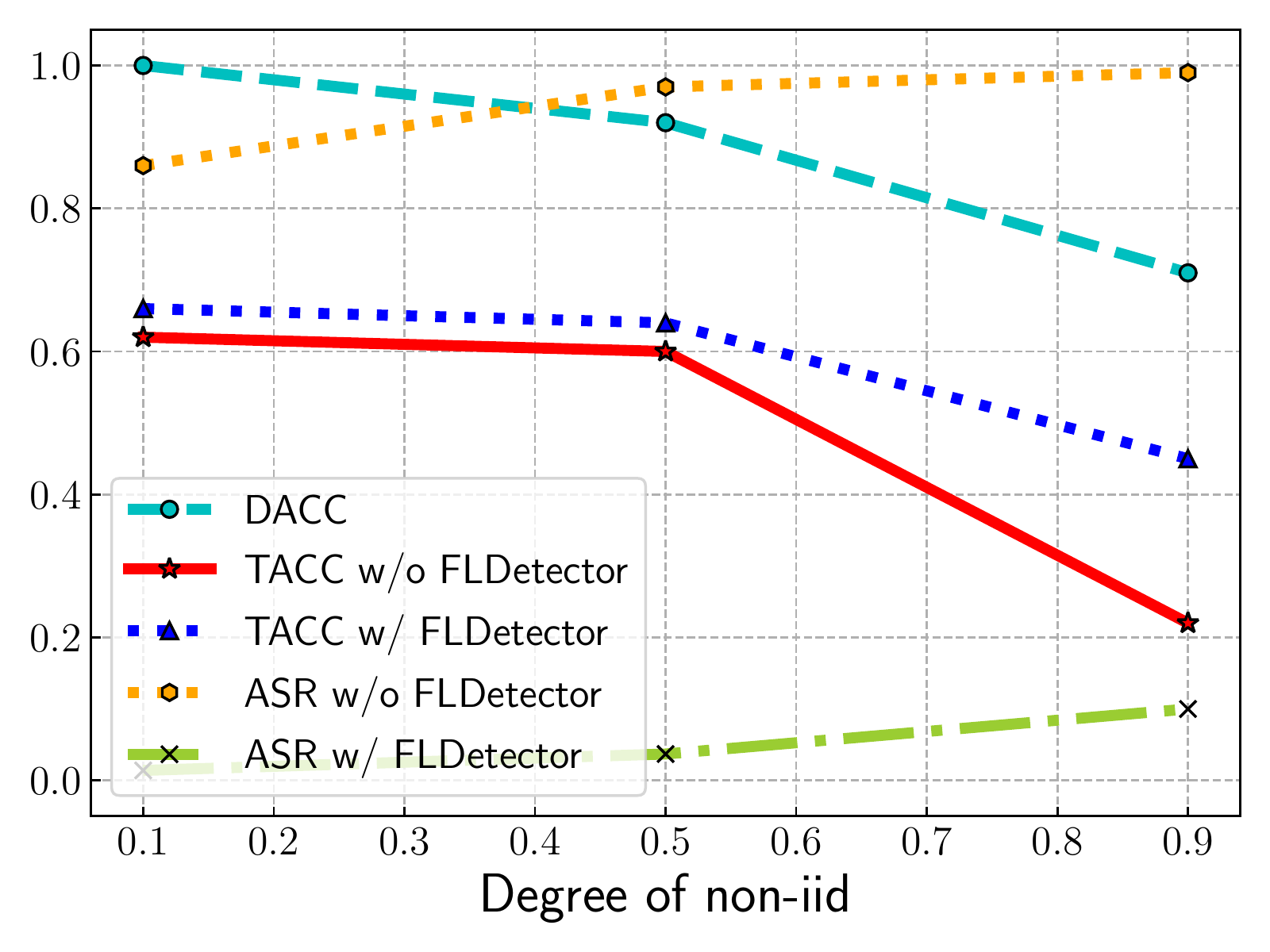}}
	    \vspace{-2mm}
	\caption{Impact of the degree of non-iid on FLDetector, where CIFAR10, Median, and 28 malicious clients are used.}
	\label{bias}
\end{figure*}

\begin{figure*}[t]
\centering
	\subfigure[Untargeted Model Poisoning Attack]{\includegraphics[width=0.24\linewidth]{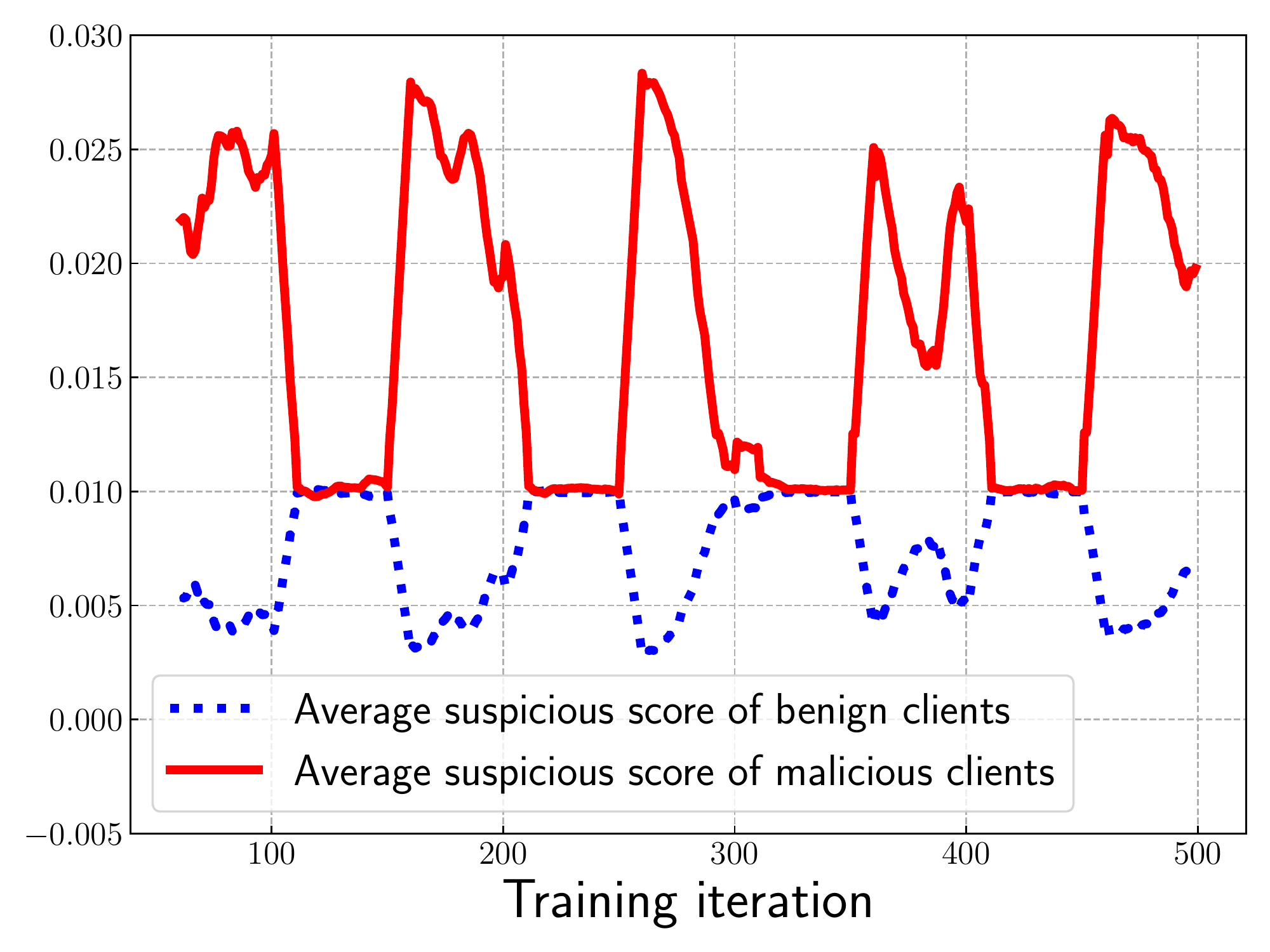}}
    \subfigure[Scaling Attack]{\includegraphics[width=0.24\linewidth]{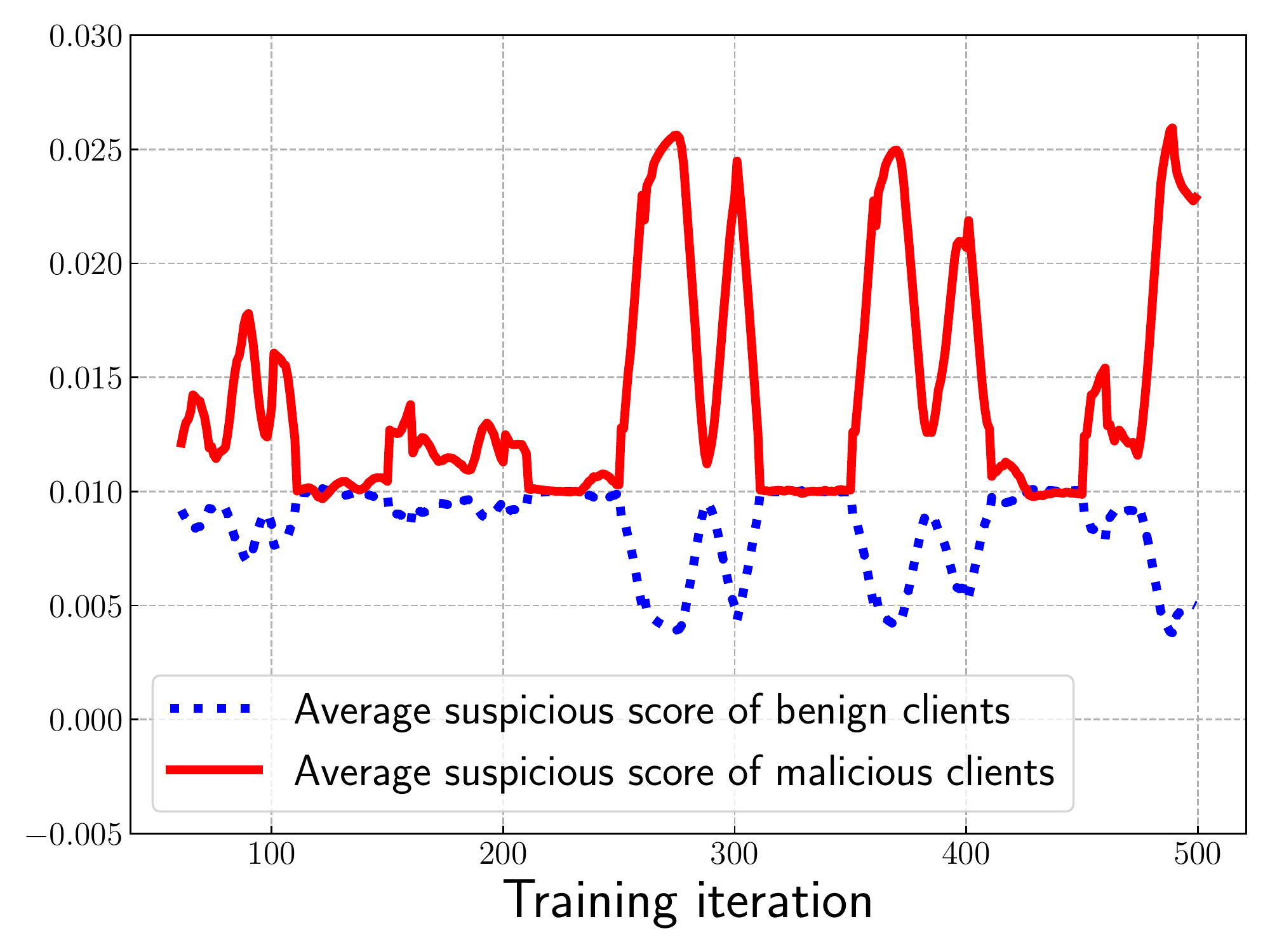}}
        \subfigure[Distributed Backdoor Attack]{\includegraphics[width=0.24\linewidth]{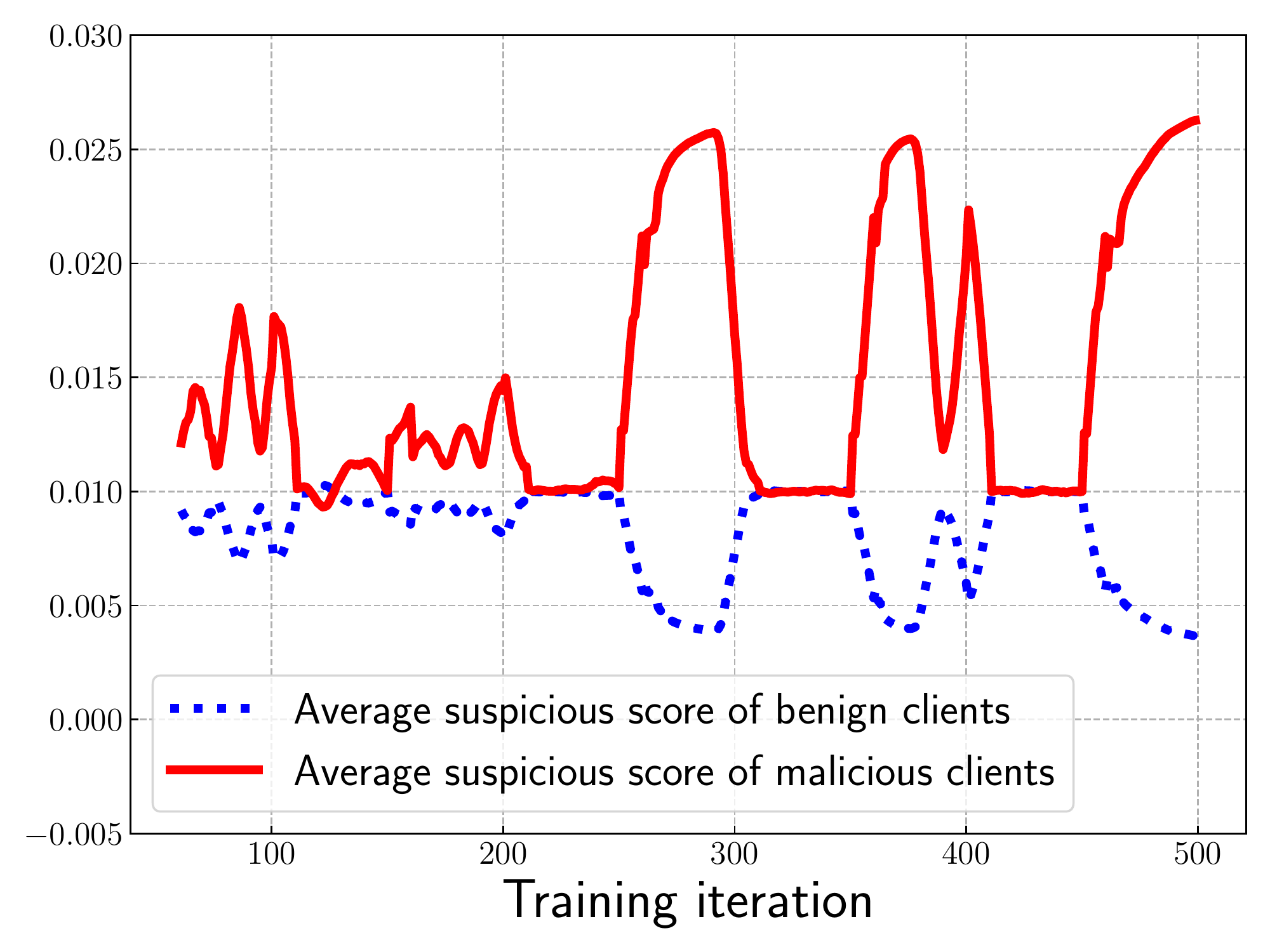}}
    \subfigure[A Little is Enough Attack]{\includegraphics[width=0.24\linewidth]{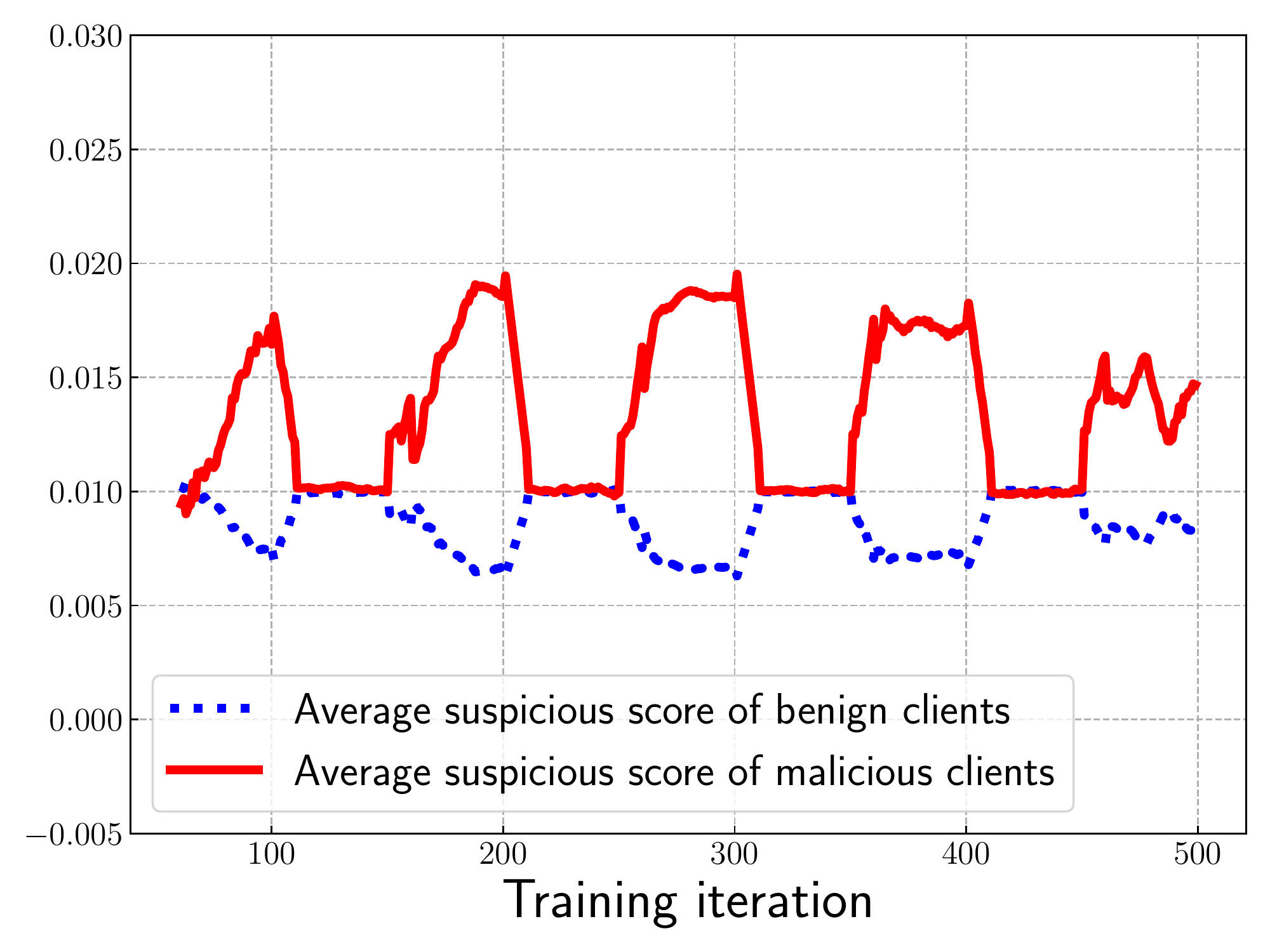}}
        \vspace{-2mm}
    \caption{Dynamics of the clients' suspicious scores when malicious clients perform attacks periodically, where  MNIST,  0.5 degree of non-iid, and 28 malicious clients are used.}
    \label{dynamic}
\end{figure*}

\myparatight{Performance of the global models} Table \ref{end-to-end} shows the TACC and ASR of the global models learnt by Median under no attacks, without FLDetector deployed, and with FLDetector deployed. Table \ref{end-to-end1} in the Appendix shows the results of other FL methods on MNIST. ``No Attack'' means the global models are learnt by Median using the remaining 72\% of benign clients; ``w/o FLDetector'' means the global models are learnt using all clients including both benign and malicious ones; and ``w/ FLDetector'' means that the server uses FLDetector to detect malicious clients, and after detecting malicious clients, the server removes them and restarts the FL training using the remaining clients. 

We observe that the global models learnt with FLDetector deployed under different attacks are as accurate as those learnt under no attacks. Moreover, the ASRs of the global models learnt with FLDetector deployed are very small. This is because after FLDetector detects and removes majority of malicious clients, Byzantine-robust FL methods can  resist the small number of malicious clients that miss detection. For instance, FLDetector misses 2 malicious clients on CIFAR10 in Median and A Little is Enough Attack, but Median is robust against them when learning the global model.

\myparatight{Impact of the number of malicious clients and degree of non-iid}  Figure \ref{client} and \ref{bias} show the impact of the number of malicious clients and the non-iid degree on FLDetector, respectively. First, we observe that the DACC of FLDetector starts to drop after the number of malicious clients is larger than some threshold or the non-iid degree is larger than some threshold, but the thresholds are attack-dependent. For instance, for the Untargeted Model Poisoning Attack, DACC of FLDetector starts to decrease after more than 30 clients are malicious, while it starts to decrease after 20 malicious clients for the A Little is Enough Attack. Second, the global models learnt with FLDetector deployed are  more accurate than the global models learnt without FLDetector deployed for different number of malicious clients and non-iid degrees. Specifically, the TACCs of the global models learnt with FLDetector deployed are larger than or comparable with those of the global models learnt without FLDetector deployed, while the ASRs of the global models learnt with FLDetector deployed are much smaller than those of the global models learnt without FLDetector. The reason is that FLDetector detects and removes (some) malicious clients.

\begin{figure*}[!t]
	\centering
    \subfigure[]{\includegraphics[width=0.24\linewidth]{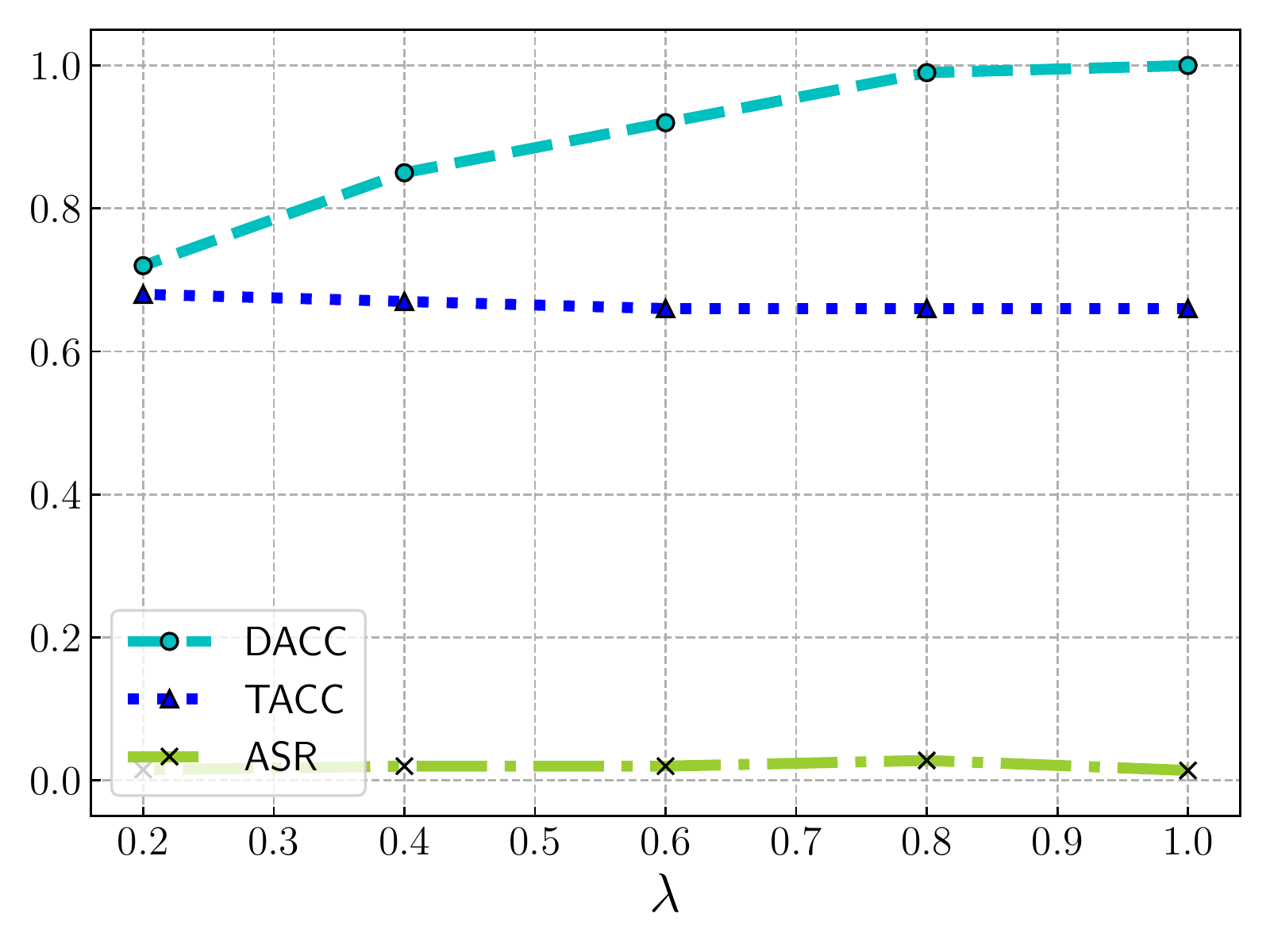}}
    \subfigure[]{\includegraphics[width=0.24\linewidth]{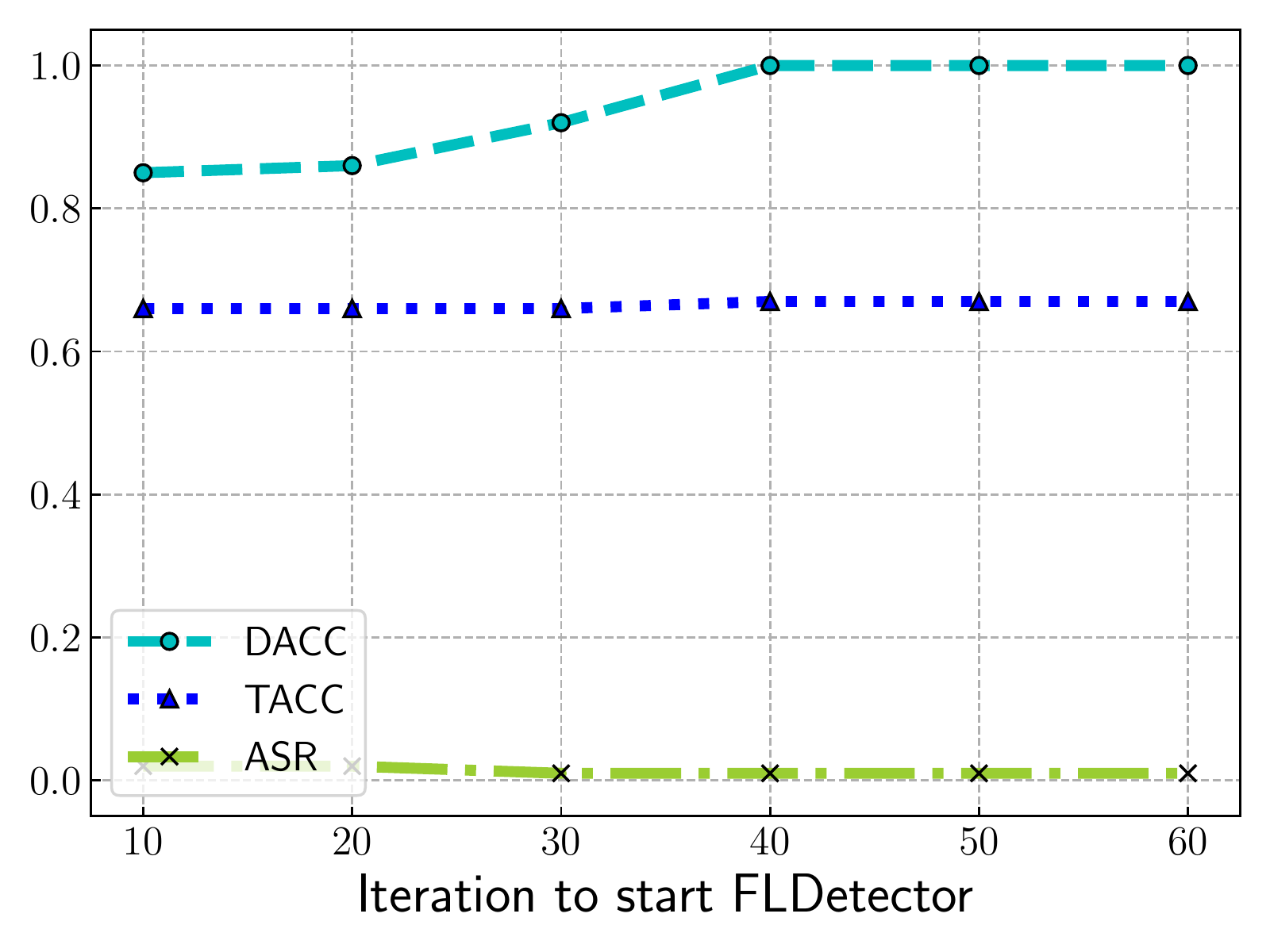}}
    \subfigure[]{\includegraphics[width=0.24\linewidth]{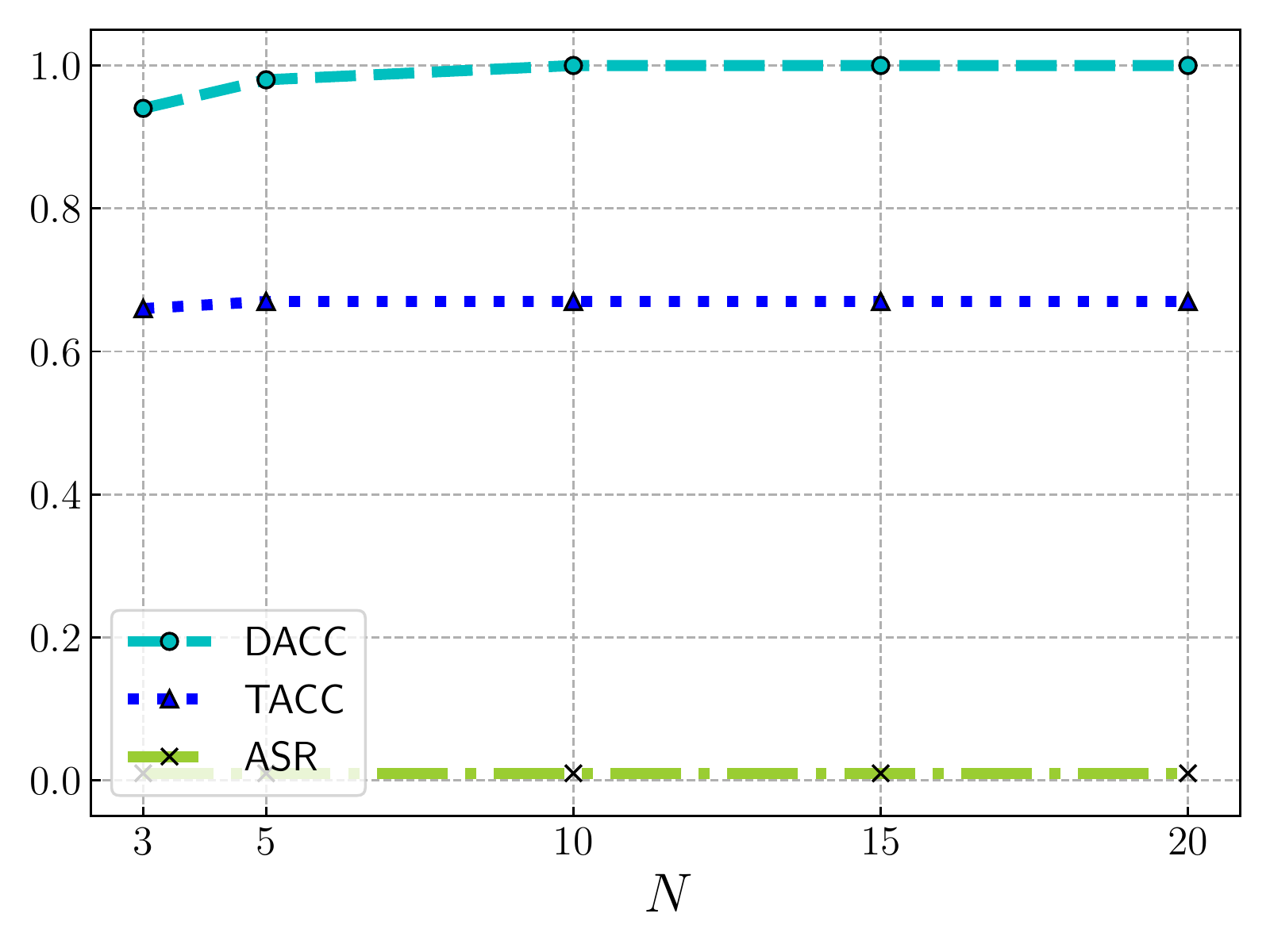}}
    \subfigure[]{\includegraphics[width=0.24\linewidth]{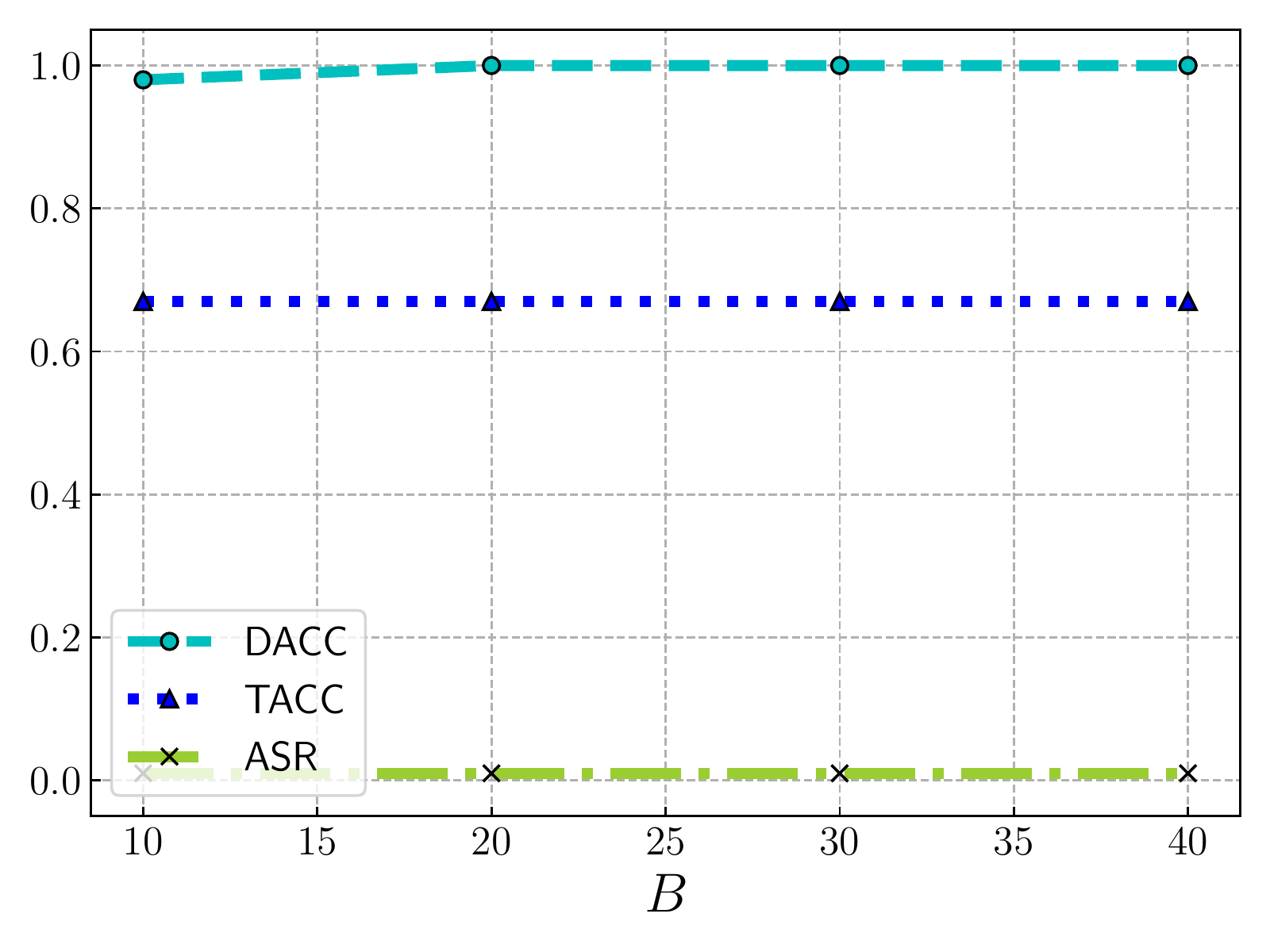}}
	\caption{(a) Adaptive attack; (b)  impact of the detection iteration; (c) impact of window size $N$; (d) impact of number of sampling $B$, where CIFAR10, Median, 0.5 degree of non-iid, 28 malicious clients, and scaling attack are used.}
	\label{adaptive_attack}
\end{figure*}

\myparatight{Dynamics of the clients' suspicious scores} Figure \ref{dynamic} shows the average suspicious scores of benign clients and malicious clients as a function of the training iteration $t$. To better show the dynamics of the suspicious scores, we assume the malicious clients perform the attacks in the first 50 iterations in every 100 iterations, starting from the 50th iteration. Note that FLDetector is ignorant of when the attack starts or ends. We observe the periodic patterns of the suspicious scores follow the attack patterns. 
Specifically, the average suspicious score of the malicious clients grows rapidly when the attack begins and  drops to be around the same as that of the benign clients when the attack stops. In the iterations where there are attacks, malicious and benign clients can be well separated based on the suspicious scores. In these experiments, FLDetector  can detect malicious clients at around 60th iteration. Note that the average suspicious score of the benign clients decreases (or increases) in the iterations where there are  attacks (or no attacks). This is because FLDetector normalizes the corresponding Euclidean distances when calculating suspicious scores.

\myparatight{Adaptive attack and impact of the detection iteration} Figure \ref{adaptive_attack}(a) shows the performance when we adapt   Scaling Attack to FLDetector. We observe that DACC drops as $\lambda$ decreases. However,  ASR is still low because the local model updates from the malicious clients are less effective while trying to evade detection. Figure \ref{adaptive_attack}(b) shows the impact of the detection iteration. Although  DACC drops slightly when FLDetector starts earlier due to the instability in the early iterations, FLDetector can still defend against  Scaling Attack by removing a majority of the malicious clients.

\myparatight{Impact of hyperparameters} Figure \ref{adaptive_attack}(c) and (d) explore the impact of hyperparameters $N$ and $B$, respectively. We observe FLDetector is robust to these hyperparameters. DACC drops slightly when $N$ is too small. This is because the suspicious scores fluctuate in a small number of rounds. In experiments, we choose $N$ = 10 and $B$ =20 as the default setting considering the trade-off between detection accuracy and computation complexity. 

\section{Conclusion and Future Work}
In this paper, we propose FLDetector, a malicious-client detection method that checks the clients' model-updates consistency. We quantify a client's model-updates consistency using the Cauchy mean value theorem and an L-BFGS algorithm. 
Our extensive evaluation on three popular benchmark datasets, four state-of-the-art attacks, and four FL methods shows that FLDetector outperforms baseline detection methods in various scenarios. Interesting future research directions include extending our method to vertical federated learning, asynchronous federated learning,  federated learning in other domains such as text and graphs, as well as efficient recovery of the global model from model poisoning attacks after  removing the detected malicious clients.  

\section*{Acknowledgements}
We thank the  reviewers for constructive comments. This work is supported by NSF under grant No. 2125977 and 2112562.


\bibliographystyle{ACM-Reference-Format}
\balance
\bibliography{main.bib}


\begin{thebibliography}{22}


\ifx \showCODEN    \undefined \def \showCODEN     #1{\unskip}     \fi
\ifx \showDOI      \undefined \def \showDOI       #1{#1}\fi
\ifx \showISBNx    \undefined \def \showISBNx     #1{\unskip}     \fi
\ifx \showISBNxiii \undefined \def \showISBNxiii  #1{\unskip}     \fi
\ifx \showISSN     \undefined \def \showISSN      #1{\unskip}     \fi
\ifx \showLCCN     \undefined \def \showLCCN      #1{\unskip}     \fi
\ifx \shownote     \undefined \def \shownote      #1{#1}          \fi
\ifx \showarticletitle \undefined \def \showarticletitle #1{#1}   \fi
\ifx \showURL      \undefined \def \showURL       {\relax}        \fi
\providecommand\bibfield[2]{#2}
\providecommand\bibinfo[2]{#2}
\providecommand\natexlab[1]{#1}
\providecommand\showeprint[2][]{arXiv:#2}

\bibitem[\protect\citeauthoryear{Bagdasaryan, Veit, Hua, Estrin, and
  Shmatikov}{Bagdasaryan et~al\mbox{.}}{2020}]%
        {bagdasaryan2020backdoor}
\bibfield{author}{\bibinfo{person}{Eugene Bagdasaryan},
  \bibinfo{person}{Andreas Veit}, \bibinfo{person}{Yiqing Hua},
  \bibinfo{person}{Deborah Estrin}, {and} \bibinfo{person}{Vitaly Shmatikov}.}
  \bibinfo{year}{2020}\natexlab{}.
\newblock \showarticletitle{How to backdoor federated learning}. In
  \bibinfo{booktitle}{\emph{AISTATS}}.
\newblock


\bibitem[\protect\citeauthoryear{Baruch, Baruch, and Goldberg}{Baruch
  et~al\mbox{.}}{2019}]%
        {baruch2019little}
\bibfield{author}{\bibinfo{person}{Gilad Baruch}, \bibinfo{person}{Moran
  Baruch}, {and} \bibinfo{person}{Yoav Goldberg}.}
  \bibinfo{year}{2019}\natexlab{}.
\newblock \showarticletitle{A Little Is Enough: Circumventing Defenses For
  Distributed Learning}. In \bibinfo{booktitle}{\emph{NeurIPS}}.
\newblock


\bibitem[\protect\citeauthoryear{Bhagoji, Chakraborty, Mittal, and
  Calo}{Bhagoji et~al\mbox{.}}{2019}]%
        {bhagoji2019analyzing}
\bibfield{author}{\bibinfo{person}{Arjun~Nitin Bhagoji},
  \bibinfo{person}{Supriyo Chakraborty}, \bibinfo{person}{Prateek Mittal},
  {and} \bibinfo{person}{Seraphin Calo}.} \bibinfo{year}{2019}\natexlab{}.
\newblock \showarticletitle{Analyzing federated learning through an adversarial
  lens}. In \bibinfo{booktitle}{\emph{ICML}}.
\newblock


\bibitem[\protect\citeauthoryear{Blanchard, El~Mhamdi, Guerraoui, and
  Stainer}{Blanchard et~al\mbox{.}}{2017}]%
        {blanchard2017machine}
\bibfield{author}{\bibinfo{person}{Peva Blanchard}, \bibinfo{person}{El~Mahdi
  El~Mhamdi}, \bibinfo{person}{Rachid Guerraoui}, {and} \bibinfo{person}{Julien
  Stainer}.} \bibinfo{year}{2017}\natexlab{}.
\newblock \showarticletitle{Machine learning with adversaries: Byzantine
  tolerant gradient descent}. In \bibinfo{booktitle}{\emph{NeurIPS}}.
\newblock


\bibitem[\protect\citeauthoryear{Byrd, Nocedal, and Schnabel}{Byrd
  et~al\mbox{.}}{1994}]%
        {byrd1994representations}
\bibfield{author}{\bibinfo{person}{Richard~H Byrd}, \bibinfo{person}{Jorge
  Nocedal}, {and} \bibinfo{person}{Robert~B Schnabel}.}
  \bibinfo{year}{1994}\natexlab{}.
\newblock \showarticletitle{Representations of quasi-Newton matrices and their
  use in limited memory methods}.
\newblock \bibinfo{journal}{\emph{Mathematical Programming}}
  (\bibinfo{year}{1994}).
\newblock


\bibitem[\protect\citeauthoryear{Caldas, Duddu, Wu, Li, Kone{\v{c}}n{\`y},
  McMahan, Smith, and Talwalkar}{Caldas et~al\mbox{.}}{2018}]%
        {caldas2018leaf}
\bibfield{author}{\bibinfo{person}{Sebastian Caldas}, \bibinfo{person}{Sai
  Meher~Karthik Duddu}, \bibinfo{person}{Peter Wu}, \bibinfo{person}{Tian Li},
  \bibinfo{person}{Jakub Kone{\v{c}}n{\`y}}, \bibinfo{person}{H~Brendan
  McMahan}, \bibinfo{person}{Virginia Smith}, {and} \bibinfo{person}{Ameet
  Talwalkar}.} \bibinfo{year}{2018}\natexlab{}.
\newblock \showarticletitle{Leaf: A benchmark for federated settings}.
\newblock \bibinfo{journal}{\emph{arXiv preprint arXiv:1812.01097}}
  (\bibinfo{year}{2018}).
\newblock


\bibitem[\protect\citeauthoryear{Cao, Fang, Liu, and Gong}{Cao
  et~al\mbox{.}}{2021a}]%
        {cao2020fltrust}
\bibfield{author}{\bibinfo{person}{Xiaoyu Cao}, \bibinfo{person}{Minghong
  Fang}, \bibinfo{person}{Jia Liu}, {and} \bibinfo{person}{Neil~Zhenqiang
  Gong}.} \bibinfo{year}{2021}\natexlab{a}.
\newblock \showarticletitle{{FLTrust}: Byzantine-robust Federated Learning via
  Trust Bootstrapping}. In \bibinfo{booktitle}{\emph{NDSS}}.
\newblock


\bibitem[\protect\citeauthoryear{Cao and Gong}{Cao and Gong}{2022}]%
        {cao2022mpaf}
\bibfield{author}{\bibinfo{person}{Xiaoyu Cao} {and}
  \bibinfo{person}{Neil~Zhenqiang Gong}.} \bibinfo{year}{2022}\natexlab{}.
\newblock \showarticletitle{{MPAF}: Model Poisoning Attacks to Federated
  Learning based on Fake Clients}. In \bibinfo{booktitle}{\emph{CVPR
  Workshops}}.
\newblock


\bibitem[\protect\citeauthoryear{Cao, Jia, and Gong}{Cao
  et~al\mbox{.}}{2021b}]%
        {cao2021provably}
\bibfield{author}{\bibinfo{person}{Xiaoyu Cao}, \bibinfo{person}{Jinyuan Jia},
  {and} \bibinfo{person}{Neil~Zhenqiang Gong}.}
  \bibinfo{year}{2021}\natexlab{b}.
\newblock \showarticletitle{Provably Secure Federated Learning against
  Malicious Clients}. In \bibinfo{booktitle}{\emph{AAAI}}.
\newblock


\bibitem[\protect\citeauthoryear{Chen, Su, and Xu}{Chen et~al\mbox{.}}{2017}]%
        {chen2017distributed}
\bibfield{author}{\bibinfo{person}{Yudong Chen}, \bibinfo{person}{Lili Su},
  {and} \bibinfo{person}{Jiaming Xu}.} \bibinfo{year}{2017}\natexlab{}.
\newblock \showarticletitle{Distributed statistical machine learning in
  adversarial settings: Byzantine gradient descent}. In
  \bibinfo{booktitle}{\emph{SIGMETRICS}}.
\newblock


\bibitem[\protect\citeauthoryear{Fang, Cao, Jia, and Gong}{Fang
  et~al\mbox{.}}{2020}]%
        {fang2020local}
\bibfield{author}{\bibinfo{person}{Minghong Fang}, \bibinfo{person}{Xiaoyu
  Cao}, \bibinfo{person}{Jinyuan Jia}, {and} \bibinfo{person}{Neil Gong}.}
  \bibinfo{year}{2020}\natexlab{}.
\newblock \showarticletitle{Local model poisoning attacks to Byzantine-robust
  federated learning}. In \bibinfo{booktitle}{\emph{USENIX Security}}.
\newblock


\bibitem[\protect\citeauthoryear{He, Zhang, Ren, and Sun}{He
  et~al\mbox{.}}{2016}]%
        {he2016deep}
\bibfield{author}{\bibinfo{person}{Kaiming He}, \bibinfo{person}{Xiangyu
  Zhang}, \bibinfo{person}{Shaoqing Ren}, {and} \bibinfo{person}{Jian Sun}.}
  \bibinfo{year}{2016}\natexlab{}.
\newblock \showarticletitle{Deep residual learning for image recognition}. In
  \bibinfo{booktitle}{\emph{CVPR}}.
\newblock


\bibitem[\protect\citeauthoryear{Krizhevsky and Hinton.}{Krizhevsky and
  Hinton.}{2009}]%
        {cifar}
\bibfield{author}{\bibinfo{person}{A. Krizhevsky} {and} \bibinfo{person}{G.
  Hinton.}} \bibinfo{year}{2009}\natexlab{}.
\newblock \bibinfo{title}{Learning multiple layers of features from tiny
  images.}
\newblock
\newblock


\bibitem[\protect\citeauthoryear{Lang}{Lang}{1968}]%
        {lang1968second}
\bibfield{author}{\bibinfo{person}{Serge Lang}.}
  \bibinfo{year}{1968}\natexlab{}.
\newblock \bibinfo{booktitle}{\emph{A second course in calculus}}.
  Vol.~\bibinfo{volume}{4197}.
\newblock \bibinfo{publisher}{Addison-Wesley Publishing Company}.
\newblock


\bibitem[\protect\citeauthoryear{LeCun.}{LeCun.}{1998}]%
        {mnist}
\bibfield{author}{\bibinfo{person}{Y. LeCun.}} \bibinfo{year}{1998}\natexlab{}.
\newblock \bibinfo{title}{The MNIST database of handwritten digits.
  http://yann. lecun. com/exdb/mnist/}.
\newblock
\newblock


\bibitem[\protect\citeauthoryear{Li, Cheng, Wang, Liu, and Chen}{Li
  et~al\mbox{.}}{2020}]%
        {li2020learning}
\bibfield{author}{\bibinfo{person}{Suyi Li}, \bibinfo{person}{Yong Cheng},
  \bibinfo{person}{Wei Wang}, \bibinfo{person}{Yang Liu}, {and}
  \bibinfo{person}{Tianjian Chen}.} \bibinfo{year}{2020}\natexlab{}.
\newblock \showarticletitle{Learning to detect malicious clients for robust
  federated learning}.
\newblock \bibinfo{journal}{\emph{arXiv preprint arXiv:2002.00211}}
  (\bibinfo{year}{2020}).
\newblock


\bibitem[\protect\citeauthoryear{McMahan, Moore, Ramage, Hampson, and
  y~Arcas}{McMahan et~al\mbox{.}}{2017}]%
        {mcmahan2017communication}
\bibfield{author}{\bibinfo{person}{Brendan McMahan}, \bibinfo{person}{Eider
  Moore}, \bibinfo{person}{Daniel Ramage}, \bibinfo{person}{Seth Hampson},
  {and} \bibinfo{person}{Blaise~Aguera y Arcas}.}
  \bibinfo{year}{2017}\natexlab{}.
\newblock \showarticletitle{Communication-efficient learning of deep networks
  from decentralized data}. In \bibinfo{booktitle}{\emph{AISTATS}}.
\newblock


\bibitem[\protect\citeauthoryear{Shejwalkar and Houmansadr}{Shejwalkar and
  Houmansadr}{2021}]%
        {shejwalkar2021manipulating}
\bibfield{author}{\bibinfo{person}{Virat Shejwalkar} {and}
  \bibinfo{person}{Amir Houmansadr}.} \bibinfo{year}{2021}\natexlab{}.
\newblock \showarticletitle{Manipulating the Byzantine: Optimizing Model
  Poisoning Attacks and Defenses for Federated Learning}. In
  \bibinfo{booktitle}{\emph{NDSS}}.
\newblock


\bibitem[\protect\citeauthoryear{Tibshirani, Walther, and Hastie}{Tibshirani
  et~al\mbox{.}}{2001}]%
        {tibshirani2001estimating}
\bibfield{author}{\bibinfo{person}{Robert Tibshirani},
  \bibinfo{person}{Guenther Walther}, {and} \bibinfo{person}{Trevor Hastie}.}
  \bibinfo{year}{2001}\natexlab{}.
\newblock \showarticletitle{Estimating the number of clusters in a data set via
  the gap statistic}.
\newblock \bibinfo{journal}{\emph{Journal of the Royal Statistical Society:
  Series B (Statistical Methodology)}} (\bibinfo{year}{2001}).
\newblock


\bibitem[\protect\citeauthoryear{Xie, Huang, Chen, and Li}{Xie
  et~al\mbox{.}}{2019}]%
        {xie2019dba}
\bibfield{author}{\bibinfo{person}{Chulin Xie}, \bibinfo{person}{Keli Huang},
  \bibinfo{person}{Pin-Yu Chen}, {and} \bibinfo{person}{Bo Li}.}
  \bibinfo{year}{2019}\natexlab{}.
\newblock \showarticletitle{Dba: Distributed backdoor attacks against federated
  learning}. In \bibinfo{booktitle}{\emph{ICLR}}.
\newblock


\bibitem[\protect\citeauthoryear{Yang, Liu, Chen, and Tong}{Yang
  et~al\mbox{.}}{2019}]%
        {yang2019federated}
\bibfield{author}{\bibinfo{person}{Qiang Yang}, \bibinfo{person}{Yang Liu},
  \bibinfo{person}{Tianjian Chen}, {and} \bibinfo{person}{Yongxin Tong}.}
  \bibinfo{year}{2019}\natexlab{}.
\newblock \showarticletitle{Federated machine learning: Concept and
  applications}.
\newblock \bibinfo{journal}{\emph{TIST}} (\bibinfo{year}{2019}).
\newblock


\bibitem[\protect\citeauthoryear{Yin, Chen, Kannan, and Bartlett}{Yin
  et~al\mbox{.}}{2018}]%
        {yin2018byzantine}
\bibfield{author}{\bibinfo{person}{Dong Yin}, \bibinfo{person}{Yudong Chen},
  \bibinfo{person}{Ramchandran Kannan}, {and} \bibinfo{person}{Peter
  Bartlett}.} \bibinfo{year}{2018}\natexlab{}.
\newblock \showarticletitle{Byzantine-robust distributed learning: Towards
  optimal statistical rates}. In \bibinfo{booktitle}{\emph{ICML}}.
\newblock


\end{thebibliography}

\appendix

\clearpage

\begin{table*}[h]
\centering
\caption{DACC, FPR, and FNR of malicious-client detection for different attacks, detection methods, and aggregation rules. The best detection results are bold for each attack. MNIST dataset, CNN global model, and 28 malicious clients are used.}
\vspace{-1mm}
\footnotesize
{
\begin{tabular}{ccp{0.42cm}<{\centering}p{0.42cm}<{\centering}p{0.42cm}<{\centering}p{0.42cm}<{\centering}p{0.42cm}<{\centering}p{0.42cm}<{\centering}p{0.42cm}<{\centering}p{0.42cm}<{\centering}p{0.42cm}<{\centering}p{0.42cm}<{\centering}p{0.42cm}<{\centering}p{0.42cm}<{\centering}}
\toprule
\multirow{2}{*}{Attack}                                                                     & \multirow{2}{*}{Detector}                                     & \multicolumn{3}{c}{FedAvg} & \multicolumn{3}{c}{Krum} &\multicolumn{3}{c}{Trimmed-Mean} & \multicolumn{3}{c}{Median}  \\ \cmidrule(lr){3-5}\cmidrule(lr){6-8}\cmidrule(lr){9-11}\cmidrule(lr){12-14}
                                                                                            &                                                              & \scriptsize DACC     & FPR     & FNR    & \scriptsize DACC      & FPR      & FNR     & \scriptsize DACC     & FPR     & FNR    & \scriptsize DACC    & FPR    & FNR    \\ \midrule
\multirow{4}{*}{\begin{tabular}[c]{@{}c@{}}Untargeted\\ Model\\ Poisoning\\ Attack\end{tabular}}                                         & VAE                               & 0.67 & 0.18 & 0.71 & 0.60 & 0.22 & 0.86 & 0.58 & 0.38 & 0.54 & 0.58 & 0.38 & 0.54   \\
& FLD-Norm                                                         & 0.68 & 0.17 & 0.71 & 0.08 & 0.89 & 1.00 & 0.28 & 1.00 & 0.00 & 0.15 & 0.79 & 1.00   \\
                                                                                            & FLD-NoHVP & 0.60 & 0.22 & 0.86 & 0.11 & 0.85 & 1.00 & 0.39 & 0.85 & 0.00 & 0.66 & 0.26 & 0.54      \\
                                                                                            & FLDetector                                                 & \textbf{0.87}       & \textbf{0.18}       & \textbf{0.00}       & \textbf{1.00}       & \textbf{0.00}       & \textbf{0.00}      & \textbf{1.00}       & \textbf{0.00}       & \textbf{0.00}  & \textbf{1.00}       & \textbf{0.00}       & \textbf{0.00}           \\ \midrule
\multirow{4}{*}{\begin{tabular}[c]{@{}c@{}}Scaling\\ Attack\end{tabular}}                                                       & VAE                                                          & 0.78    & 0.31    & 0.00   & 0.97   & 0.00   & 0.11 & 0.76     & 0.00     & 0.86    & 0.75    & 0.00    & 0.89      \\& FLD-Norm                                                         & \textbf{0.97}    & \textbf{0.00}    & \textbf{0.11} & 0.97   & 0.00   & 0.11   & 0.92     & 0.11     & 0.00    & \textbf{1.00}    & \textbf{0.00}    &\textbf{0.00}     \\
                                                                                            & FLD-NoHVP & 0.62    & 0.21    & 0.82  & 0.59   & 0.40   & 0.43  & 0.90     & 0.10     & 0.11    & 0.83    & 0.21    & 0.07     \\
                                                                                            & FLDetector                                                   & 0.81    & 0.22    & 0.11  & \textbf{0.98}   & \textbf{0.00}   & \textbf{0.07}  & \textbf{1.00}       & \textbf{0.00}       & \textbf{0.00}    & \textbf{1.00}       & \textbf{0.00}       & \textbf{0.00}     \\ \midrule
\multirow{4}{*}{\begin{tabular}[c]{@{}c@{}}Distributed\\Backdoor\\Attack\end{tabular}}                                                   & VAE                 & 0.89      & 0.15       & 0.00    & \textbf{0.97}       & \textbf{0.00 }     & \textbf{0.11}      & 0.79        & 0.00        & 0.75        & 0.81       & 0.06       & 0.54                \\
& FLD-Norm          & \textbf{0.91}        & \textbf{0.08}       & \textbf{0.11}      & 0.75       & 0.26      & 0.21  & 0.90        & 0.14        & 0.00        & 0.93       & 0.10       & 0.00       \\
                                                                            & FLD-NoHVP   &0.62       & 0.21       & 0.82       &  0.82       & 0.21      & 0.11   & \textbf{1.00}       & \textbf{0.00 }      & \textbf{0.00}      & 0.93       &0.10 &0.00                \\
                                                                                            & FLDetector                                    &0.86 &0.15&0.11  & \textbf{0.97}       & \textbf{0.00}        & \textbf{1.00}     & \textbf{1.00}       & \textbf{0.00 }      & \textbf{0.00}    & \textbf{1.00}           & \textbf{0.00 }     & \textbf{0.00}        \\ \midrule
\multirow{4}{*}{\begin{tabular}[c]{@{}c@{}}A Little\\is Enough\\Attack\end{tabular}}
&VAE                                                         & 0.80     & 0.28      & 0.00   & \textbf{1.00}       & \textbf{0.00}       & \textbf{0.00}     & \textbf{1.00}       & \textbf{0.00}       & \textbf{0.00}  & \textbf{1.00}       & \textbf{0.00}       & \textbf{0.00}   \\  & FLD-Norm  & 0.00        & 1.00       & 1.00  & 0.12       & 0.83      & 1.00      & 0.00        & 1.00        & 1.00        & 0.09       & 0.86       & 1.00       \\                                          & FLD-NoHVP & 0.65       & 0.10       & 1.00   & 0.02       & 0.97      & 1.00     & \textbf{1.00}       & \textbf{0.00}       & \textbf{0.00}       & 0.75       & 0.35       & 0.00            \\
& FLDetector                                                   & \textbf{1.00}       & \textbf{0.00}       & \textbf{0.00}       & \textbf{1.00}       & \textbf{0.00}       & \textbf{0.00}        & \textbf{1.00}       & \textbf{0.00}       & \textbf{0.00}       & \textbf{1.00}       & \textbf{0.00}       & \textbf{0.00}       \\ \bottomrule
\end{tabular}}

\label{detection}
\end{table*}

\begin{table*}[h]
\centering
\caption{DACC, FPR, and FNR of malicious-client detection for different attacks, detection methods, and aggregation rules. The best detection results are bold for each attack. CIFAR10 dataset, ResNet20 global model, and 28 malicious clients are used.}
\vspace{-1mm}
\footnotesize
\begin{tabular}{ccp{0.42cm}<{\centering}p{0.42cm}<{\centering}p{0.42cm}<{\centering}p{0.42cm}<{\centering}p{0.42cm}<{\centering}p{0.42cm}<{\centering}p{0.42cm}<{\centering}p{0.42cm}<{\centering}p{0.42cm}<{\centering}p{0.42cm}<{\centering}p{0.42cm}<{\centering}p{0.42cm}<{\centering}}
\toprule
\multirow{2}{*}{Attack}                                                                     & \multirow{2}{*}{Detector}                                     & \multicolumn{3}{c}{FedAvg} & \multicolumn{3}{c}{Krum}& \multicolumn{3}{c}{Trimmed-Mean} & \multicolumn{3}{c}{Median}  \\ \cmidrule(lr){3-5}\cmidrule(lr){6-8}\cmidrule(lr){9-11}\cmidrule(lr){12-14}
                                                                                            &                                                              & \scriptsize DACC     & FPR     & FNR    & \scriptsize DACC      & FPR      & FNR     & \scriptsize DACC     & FPR     & FNR    & \scriptsize DACC    & FPR    & FNR    \\ \midrule
\multirow{4}{*}{\begin{tabular}[c]{@{}c@{}}Untargeted\\ Model\\ Poisoning\\ Attack\end{tabular}}                                           & VAE                               & 0.28      & 1.00       & 0.00     & 0.61       & 0.42      & 0.32    & 0.52        & 0.33        & 0.86        & 0.46       & 0.40       & 0.89       \\& FLD-Norm                                                         & 0.72        & 0.00       & 1.00    & 0.07       & 0.90      & 1.00   & 0.00        & 1.00        & 1.00        & 0.00       & 1.00       & 1.00          \\
                                                                                            & FLD-NoHVP & 0.53       & 0.40       & 0.64  & 0.85       & 0.21      & 0.00       & 0.48        & 0.51        & 0.54        & 0.98       & 0.00       & 0.07         \\
                                                                                            & FLDetector                                                 & \textbf{0.93}       & \textbf{0.10}       & \textbf{0.00}   & \textbf{0.97}       & \textbf{0.04}      & \textbf{0.00}    & \textbf{1.00}        & \textbf{0.00}        & \textbf{0.00}        & \textbf{1.00}       & \textbf{0.00}       & \textbf{0.00}                      \\ \midrule
\multirow{4}{*}{\begin{tabular}[c]{@{}c@{}}Scaling\\ Attack\end{tabular}}                                                           & VAE                                                           & 0.24      & 0.71       & 0.89    & 0.50       & 0.31      & 1.00    & 0.48        & 0.38        & 0.89        & 0.75       & 0.00       & 0.89           \\& FLD-Norm                                                          & \textbf{0.96}        & \textbf{0.00}       & \textbf{0.14}    & 0.98       & 0.00      & 0.07     & 0.96        & 0.01        & 0.11        & \textbf{1.00}       & \textbf{0.00}       & \textbf{0.00 }       \\
                            & FLD-NoHVP & 0.86       & 0.14       & 0.14    & 1.00       & 0.00      & 0.00   & 1.00        & 0.00        & 0.00        & 0.97       & 0.00       & 0.11        \\
                                                                                            & FLDetector                                                   & 0.88       & 0.11       & 0.14       & \textbf{1.00}        & \textbf{0.00}        & \textbf{0.00 }       & \textbf{1.00}       & \textbf{0.00}       & \textbf{0.00}       & \textbf{1.00}       & \textbf{0.00}      & \textbf{0.00}  \\ \midrule
\multirow{4}{*}{\begin{tabular}[c]{@{}c@{}}Distributed\\Backdoor\\Attack\end{tabular}}                                                                     & VAE                  & 0.27      & 0.74       & 0.93    & 0.53       & 0.26      & 1.00    & 0.55        & 0.33        & 0.71        & 0.76       & 0.00       & 0.86       \\
& FLD-Norm         & \textbf{0.91}        & \textbf{0.07}       & \textbf{0.14}    & 0.85       & 0.14      & 0.18   & 0.92        & 0.07        & 0.11        & 0.96       & 0.01       & 0.11     \\
                                                                                            & FLD-NoHVP & 0.84       & 0.14       & 0.21   & 1.00       & 0.00      & 0.00       & 1.00        & 0.00        & 0.00        & 1.00       & 0.00       & 0.00            \\
                                                                                            & FLDetector                                      & 0.89       & 0.11       & 0.11       & \textbf{1.00}        & \textbf{0.00}        & \textbf{0.00}        & \textbf{1.00}       & \textbf{0.00}       & \textbf{0.00}       & \textbf{1.00}       & \textbf{0.00}      & \textbf{0.00}          \\ \midrule
\multirow{4}{*}{\begin{tabular}[c]{@{}c@{}}A Little\\is Enough\\Attack\end{tabular}}         
&VAE                                                         & \textbf{0.81}      & \textbf{0.13}       & \textbf{0.36}    & 0.61       & 0.49      & 0.14    & 0.84        & 0.14        & 0.21       & 0.85       & 0.10 & 0.29         \\
& FLD-Norm  & 0.33        & 0.69       & 0.61    & 0.77       & 0.22      & 0.25      & 0.45        & 0.69        & 0.18        & 0.47       & 0.67       & 0.18         \\ 
                                                                                            & FLD-NoHVP & 0.72       & 0.24       & 0.39  & 0.85       & 0.14      & 0.18      & 0.80        & 0.17        & 0.29        & 0.81       & 0.15       & 0.32        \\
& FLDetector                                                   & 0.80       & 0.14      & 0.36   & \textbf{0.92}       & \textbf{0.11}      & \textbf{0.00}     & \textbf{0.89}        & \textbf{0.13}        & \textbf{0.07}        & \textbf{0.87}       & \textbf{0.15}       & \textbf{0.07}   \\ \bottomrule
\end{tabular}
\label{detection1}
\end{table*}

\begin{table*}[h]
\footnotesize
\centering

\caption{TACC and ASR of the global models learnt by different FL methods on MNIST. The results for the targeted model poisoning attacks are in the form of “TACC / ASR ($\%$)”. 28$\%$ malicious clients are used.  
The ASR on FedAvg with FLDetector is still high because FedAvg is not Byzantine-robust and can be backdoored by even a single malicious client.} 

\vspace{-1mm}

\begin{tabular}{llccc}
\toprule
FL Method                 & Attack                & No Attack &  \begin{tabular}[c]{@{}c@{}} w/o FLDetector\end{tabular} & \begin{tabular}[c]{@{}c@{}} w/ FLDetector\end{tabular}\\ \midrule
\multirow{4}{*}{FedAvg}   &Untargeted Model Poisoning Attack   & 98.4 & 10.1 & 98.3                 \\
                         & Scaling Attack                & 98.4 & 98.5/99.8 & 98.2/99.6             \\
                         & Distributed Backdoor Attack   &98.4 & 98.4/99.9 & 98.1/99.5            \\
                         & A Little is Enough Attack        & 98.4 & 97.9/99.9 & 98.2/0.3            \\ \midrule
\multirow{4}{*}{Krum} &Untargeted Model Poisoning Attack   & 93.5 & 11.2 & 92.8                \\
                         & Scaling Attack                & 93.5 & 94.3/0.9 & 93.2/0.7            \\
                         & Distributed Backdoor Attack    & 93.5 & 94.4/0.8 & 93.1/0.8             \\
                         & A Little is Enough Attack            & 93.5 & 94.4/99.6 & 93.4/0.6             \\ \midrule
\multirow{4}{*}{Trimmed-Mean} &Untargeted Model Poisoning Attack   & 97.6 & 63.9 & 97.5               \\
                         & Scaling Attack               & 97.6 & 97.5/0.6 & 97.5/0.5             \\
                         & Distributed Backdoor Attack    & 97.6 & 97.5/0.5 & 97.4/0.4            \\
                         & A Little is Enough Attack          & 97.6 & 97.8/100.0 & 97.5/0.4             \\ \midrule
\multirow{4}{*}{Median} &Untargeted Model Poisoning Attack  &97.6    & 69.5                  & 97.4               \\
                         & Scaling Attack                &97.6     & 97.6/0.5              & 97.6/0.5             \\
                         & Distributed Backdoor Attack    &97.6     & 97.4/0.5              & 97.5/0.4             \\
                         & A Little is Enough Attack           &97.6    & 97.8/100.0            & 97.9/0.3             \\ \bottomrule
\end{tabular}
\label{end-to-end1}
\end{table*}

\section{Proof of Theorem 1}
\begin{lem}
For any $t$ and any vector $z$, the following inequality related to the estimated Hessian $\hat{H}^t$ holds:
\begin{equation}
    z^T \hat{H}^t z \le (N+1)L\|z\|^2,
\end{equation}
where N is the window size and L is from Assumption 1.
\end{lem}
\begin{proof}
By following Equation 1.2 and 1.3 in \cite{byrd1994representations}, the Quasi-Hessian update can be written as:
\begin{equation}
    B_{t-m+1} = B_{t-m} - \frac{B_{t-m}\Delta w_{t-m}\Delta w_{t-m}^T B_{t-m}}{\Delta w_{t-m}^T B_{t-m} \Delta w_{t-m}}+\frac{\Delta g_{t-m}\Delta g_{t-m}^T}{\Delta g_{t-m}^T\Delta w_{t-m}},
    \label{quasi_hessian}
\end{equation}
where the initialized matrix $B_{t-N} = \Delta g_{t-N}^T \Delta w_{t-N}/ \Delta w_{t-N}^T \Delta w_{t-N} \mathbf{I}$ and $m \in \{1, 2, \cdots, N\}$. The final estimated Hessian $\hat{H}^t$ = $B_t$.\par
Based on Equation \ref{quasi_hessian}, we derive an upper bound for $z^T \hat{H}^t z$:
\begin{align}
    z^T B_{t-m+1}z &= z^T B_{t-m} z - \frac{z^T B_{t-m}\Delta w_{t-m}\Delta w_{t-m}^T B_{t-m} z}{\Delta w_{t-m}^T B_{t-m} \Delta w_{t-m}}\\&+\frac{z^T \Delta g_{t-m}\Delta g_{t-m}^T z}{\Delta g_{t-m}^T\Delta w_{t-m}}\\
    &\le z^T B_{t-m} z +\frac{z^T \Delta g_{t-m}\Delta g_{t-m}^T z}{\Delta g_{t-m}^T\Delta w_{t-m}}\\
    &= z^T B_{t-m} z +\frac{z^T H_{t-m}\Delta w_{t-m}\Delta w_{t-m}^T H_{t-m}z}{\Delta w_{t-m}^T H_{t-m} \Delta w_{t-m}}\\
    &\le z^T B_{t-m} z +\frac{z^T H_{t-m}z\Delta w_{t-m}\Delta w_{t-m}^T H_{t-m}}{\Delta w_{t-m}^T H_{t-m} \Delta w_{t-m}}\\
    &=  z^T B_{t-m} z+ z^T H_{t-m}z\\
    &\le  z^T B_{t-m} z+ L\|z\|^2
\end{align}

The first inequality uses the fact that $z^T B_{t-m}\Delta w_{t-m}\Delta w_{t-m}^T B_{t-m} z = (z^T B_{t-m}\Delta w_{t-m})^2 \ge 0$ and $\Delta w_{t-m}^T B_{t-m} \Delta w_{t-m} \ge 0$ due to the positive definiteness of $B_{t-m}$. The second inequality uses the Cauchy-Schwarz inequality.\par
By applying the formula above recursively, we have $z^T \hat{H}^t z = z^T B_t z \le (N+1)L\|z\|^2$.
\end{proof}

Next, we prove Theorem 1. Our idea is to bound the difference $d_i$ between predicted model updates and the received ones from benign clients in each iteration. For $i \in \mathcal{B}$ and $j \in \mathcal{M}$, we have:
\begin{align}
    &\mathbb{E}~d_j - \mathbb{E}~d_i\\ &= \mathbb{E} \|g_j^{t-1} + \hat{H}^t(w_t-w_{t-1}) - g_j^t\| - \mathbb{E} \|g_i^{t-1} + \hat{H}^t(w_t-w_{t-1}) - g_i^t\| \\
    & = \mathbb{E}_{D_j \sim D} \|\nabla f (D_j, w_{t-1})+\nabla f (D_j, w_{t})+\hat{H}^t(w_t-w_{t-1}) \| \\ 
    &- \mathbb{E}_{D_i \sim D} \|\nabla f (D_i, w_{t-1})-\nabla f (D_i, w_{t})+\hat{H}^t(w_t-w_{t-1}) \|\\
    & \ge \mathbb{E}_{D_j \sim D} 2\|\nabla f (D_j, w_{t-1})\| - L \|w_t-w_{t-1}\| - \|\hat{H}^t(w_t-w_{t-1})\|\\
    & - \mathbb{E}_{D_i \sim D} (\|\nabla f (D_i, w_{t-1})-\nabla f (D_i, w_{t})\|+\|\hat{H}^t(w_t-w_{t-1}) \|)\\
    & \ge \mathbb{E}_{D_j \sim D} 2\|\nabla f (D_j, w_{t-1})\| -2(L \|w_t-w_{t-1}\| + \|\hat{H}^t(w_t-w_{t-1})\|)\\
    & \ge \mathbb{E}_{D_j \sim D} 2\|\nabla f (D_j, w_{t-1})\| - 2(N+2)L\|(w_t-w_{t-1})\|\\
    & = \mathbb{E}_{D_j \sim D} 2\|\nabla f (D_j, w_{t-1})\| - 2(N+2)L\alpha\mathbb{E}_{D_i \sim D}\|\nabla f (D_j, w_{t-1})\|\\
    & = (2-2(N+2)L\alpha) \mathbb{E}_{D_j \sim D} \|\nabla f (D_j, w_{t-1})\| \\
    & \ge 0,
\end{align}
where the first inequality uses the Triangle inequality, the second inequality uses Assumption 1, and the third inequality uses Lemma 1. According to the definition of suspicious scores ($s_i^t=\frac{1}{N}\sum_{r=0}^{N-1} \hat{d}_i^{t-r}$), we have $\mathbb{E}~(s_i^t) < \mathbb{E}~(s_j^t)$.

\vfill\eject
\end{document}